\documentclass{aa}  

\usepackage{amsmath}
\usepackage{graphicx}
\usepackage{changepage}
\usepackage{listings}
\usepackage{siunitx}
\usepackage{natbib}
\usepackage{hyperref}
\usepackage{multirow}
\usepackage{float}
\usepackage{placeins}

\usepackage{lastpage}
\usepackage{txfonts}
\usepackage{xcolor}
\newcommand{\astext}[1]{{\textcolor{black}{#1}}}
\newcommand{\gnstext}[1]{{\textcolor{black}{#1}}}

\begin{document}

   \title{Minimum and maximum mass-luminosity relations \\ for stripped stars}


\author{{Gautham N. Sabhahit\inst{\ref{AOP}}}
    \and 
       {Jorick S. Vink\inst{\ref{AOP}}}
    \and
      {Andreas\,A.\,C. Sander\inst{\ref{ARI},\ref{IWR}}}
    \and
      {Varsha Ramachandran\inst{\ref{ARI}}}
    }

\institute{
   {Armagh Observatory and Planetarium, College Hill, Armagh BT61 9DG, N. Ireland\label{AOP}}
   \and
   {Zentrum f{\"u}r Astronomie der Universit{\"a}t Heidelberg, Astronomisches Rechen-Institut, M{\"o}nchhofstr. 12-14, 69120 Heidelberg, Germany\label{ARI}}
    \and
   {Interdisziplin{\"a}res Zentrum f{\"u}r Wissenschaftliches Rechnen, Universit{\"a}t Heidelberg, Im Neuenheimer Feld 225, 69120 Heidelberg, Germany\label{IWR}}\\
   \email{gauthamns96@gmail.com}
             }


 
  \abstract
   {Envelope stripping, whether through single-star wind mass loss or binary mass transfer, is a key evolutionary pathway for the formation of classical Wolf-Rayet stars and lower-mass stripped helium (He) stars. However, to study the evolution of these objects into black holes, neutron stars, and stripped-envelope supernovae, we need appropriate input models for the core-He burning phase without relying on the uncertain evolution into this evolved phase. Reliable mass-luminosity relations (MLRs) for He stars are needed for stellar wind and evolution studies, but the MLRs currently in literature are either for fully-stripped or chemically homogeneous stars, neither of which reflect the important and recently also observationally confirmed stage of partial stripping. We alleviate this drawback by computing sets of MESA synthetic structure models with partially-stripped chemical profiles, consisting of a pure-He core and a hydrogen (H)-depleted envelope with an H/He chemical gradient left behind from the receding convective core during the main sequence. As the H-profile slope increases from 0 (full chemical homogeneity) to $\infty$ (pure-He stars) in our synthetic models, we find the luminosity to initially increase before eventually decreasing. The maximum luminosity for a given mass is reached for an intermediate H-profile slope corresponding to a partially-stripped structure, exceeding even the values documented for pure-He stars, primarily due to the H shell disproportionately dominating the total luminosity budget. We also provide convenient mass-luminosity fit relations to predict the minimum, maximum, and pure-He luminosities for a given mass -- and vice versa -- while accounting for structures achievable through partial stripping. We also explore the impact of the higher luminosity on the wind properties of partially-stripped configurations using hydrodynamically consistent atmosphere models.  
   }

   \keywords{Stars: fundamental parameters -- Stars: massive -- Stars: mass loss -- Stars: evolution -- stars: Wolf-Rayet -- (Stars:) binaries: general
               }

   \maketitle
%

\section{Introduction}
\label{sec: Introduction}

Massive stars dominate the upper regions of the Hertzsprung-Russell diagram (HRD) and, depending on their initial mass, can be connected to a range of evolutionary pathways --- from hot O stars on the main sequence (MS) to cool supergiants, or, if they lose their hydrogen (H) envelopes, to the hot Wolf-Rayet (WR) phase with temperatures of several hundred thousand kelvin \citep[e.g., see evolutionary calculations by][]{Brott2011, Ekstrom2012, Kohler2015, Mist2016}. Adding to this complexity, nearly all massive stars are members of binary systems, with about half found in close binaries expected to interact during their lifetimes \citep[][]{Sana2013}. Such interactions, including mass transfer and mergers, are predicted to give rise to a diverse population of post-interaction objects, scattered across the upper HRD \citep{Paczynski1967, Podsiadlowski1992, Petrovic2005, Eldridge2008, Schneider2015, Laplace2020, Renzo2021, Klencki2022}. 

An open question in stellar astrophysics is the formation scenario of classical WR (cWR) stars, which are spectroscopically defined by their strong and broad emission lines, as well as their lower-mass stripped-He counterparts that lack such emission features. These stars have very little to no H detected in their atmospheres, indicating that they are evolved objects that have undergone significant prior envelope stripping. While we have a relatively good understanding of the redward evolution of massive stars from their zero-age MS (ZAMS) -- including both the extension of the MS width and the rapid expansion across the Hertzsprung gap (HG) into red supergiant (RSG) proportions -- the key question remains: what drives these stars to evolve bluewards again? Whether this is achieved through mass loss and mixing in single stars \citep{Meynet2003} or through binary interaction where mass transfer fully or partially strips the H-rich envelope \citep{Podsiadlowski1992, Gilkis2019}, is still an open question. 

In recent years, there has been growing interest in the stripped star evolution into the cWR phase and their lower mass counterparts. For example, efforts have been made in the observational front, with spectroscopic analysis of fully- and partially-stripped stars in the Magellanic Clouds (MCs) to determine their stellar properties \citep{Gotberg2023, Drout2023, Ramachandran2023, Ramachandran2024}. Multiple partially-stripped stars have been observed with surface H mass fractions lower than the H ZAMS value of roughly 0.7 and positions on the HRD that are cooler than the helium (He) ZAMS and, at times, even cooler than the H ZAMS. Detailed binary evolutionary models at sub-Galactic metallicity have been explored to explain such surface properties as the lower metal content in the Clouds compared to the Galaxy leads to less efficient wind mass loss \citep{Abbott1982, Vink2001, Kudritzki2002} and binary stripping \citep{Gotberg2017, Klencki2022, Dutta_Klencki2024}, making partial stripping relatively more common.

However, the evolution bluewards is affected by uncertainties in single and binary massive star evolution during the prior redward evolution, particularly those related to mass loss through winds or binary stripping, the stability and efficiency of mass transfer, and the mixing efficiencies associated with the various instabilities (for example, semiconvection) implemented in numerical codes. Whether a star fully strips and exposes the pure-He core or undergoes only partial stripping, remaining cooler than the He ZAMS, is influenced by these uncertainties in its prior evolution. These modelling uncertainties affect the final masses and chemical profiles of the stars, ultimately shaping their end fates as stripped-envelope supernovae (SNe), black holes, or neutron stars. 

Given the uncertainties involved in evolutionary modelling, a possible way forward is reliable mass-luminosity relations (MLRs) that put constraints on the maximum and minimum possible masses (or luminosities) for a given luminosity (or mass) of a partially-stripped or fully-stripped star. There are already MLRs in existing literature that predict the possible mass ranges for a given luminosity and surface H abundance \citep{Graf2011}. However, such MLRs have been devised either for the case of full chemical homogeneity or full stripping, that is, with pure-He configurations. Here, we extend on the work of \citet{Graf2011}, by also including structure configurations obtained from partial envelope stripping.

The reason for revisiting the MLRs in the context of partial envelope stripping in this work is the resulting non-intuitive structure. \gnstext{A typical \textit{partially-stripped structure} consists of a He-burning core surrounded by a low-mass, H-depleted envelope, where H burns in a shell. This is the most common configuration, with stars spending up to a few $10^5$ years in this phase \citep{Dutta_Klencki2024}, whereas post-He-burning giants with He and/or H shells that re-expand after core-He exhaustion are significantly rarer due to their shorter thermal timescale evolution. In this work, we primarily focus on structures with the typical He-burning core and H-burning shell configuration.}

\gnstext{Strong single-star wind mass loss could result in such structures. They can also result from a binary channel via a post-MS (or case B) mass transfer phase, followed by detachment, resulting in the donor having a large(r) He core compared to the H envelope. Envelope stripping can also occur from a mass transfer phase during the MS (or case A). However, the post-detachment donor will still be core-H-burning and will not have a H-burning shell or the non-intuitive structure we are interested in here.}

What is particularly intriguing about the partially-stripped structure is the often disproportionate contribution of the H shell to the total luminosity. Despite the He core occupying a large fraction of the total mass, the H shell dominates the total luminosity. This can lead to a breakdown of simple homology relations. For a given mass, the pure-He configuration (i.e., the configuration with the highest mean molecular weight $\mu$) might not have the highest luminosity. Instead, a partially-stripped structure can have a higher luminosity due to the H shell dominating over the He core. 

Partially-stripped stars also pose an interesting problem, as they overlap with other class of objects in the HRD space. For example, there is ongoing debate regarding the evolutionary status of B supergiants -- whether they are still core-H-burning objects or have already evolved beyond the main sequence \citep{Vink2010} -- since the width of the MS depends sensitively on the efficiency of core boundary mixing \citep{Maeder1976, Matraka1982, Kippenhahn2013}. To murky the waters, these partially-stripped stars could be hiding in the same region of the HRD as MS objects and supergiants evolving redwards from the H ZAMS \citep{Pauli2022}. Now, despite having the same exact luminosity and effective temperature, the two evolutionary states -- redward-evolving  supergiants, and blueward-evolving partially-stripped stars -- will have very different masses and look spectroscopically different, especially in their $\mathrm{log}\,g$ diagnostics \citep{Bernini-Peron2024, Ramachandran2024}. Partially-stripped stars can have a lower mass and produce the same total luminosity as their less chemically evolved supergiant counterparts due to their more evolved nature.

Either having lower mass compared to their B supergiant counterparts of the same luminosity, or being more luminous compared to their same-mass pure-He counterparts, they mean the same thing: partially-stripped systems have a higher ratio of radiative acceleration to gravity (known as the Eddington parameter). This will impact both the internal structure as well as wind properties of these objects. 1D structure models close to the Eddington limit predict the development of a very low-density, inflated morphology \citep{Ishi1999, Petrovic2005, Graf2012, Sanyal2015, Sabhahit2025}, while radiation-driven wind simulations in different parameter regimes show mass-loss scaling with the Eddington parameter \citep{Vink2001, GH2008,Vink2011, Graf2011, Sander2020b}.

Given the recent flurry of interest in the evolutionary nature of partially-stripped stars, along with their peculiar structure and luminosity profile, a re-investigation of MLRs is warranted. The paper is organized as follows. In $\mathrm{Sect.}\,\ref{sec: methodology}$, we describe our choice of stellar structure code to build our synthetic model grid, which consists of a He-rich core and a low-mass H shell with a realistic chemical profile. We explore a range of H/He slopes left behind by the receding convective core during the MS, informed by detailed evolutionary models. In $\mathrm{Sect.}\,\ref{sec: MLR}$, we present convenient MLR fits capable of predicting the minimum, maximum and pure-He luminosities for a given mass and surface H mass fraction. In $\mathrm{Sect.}\,\ref{sec: powr_wind_pss}$, we present hydrodynamically consistent wind models using the PoWR atmosphere code. We discuss implications of our results in the context of recently observed partially-stripped stars in the MCs in $\mathrm{Sect.}\,\ref{sec: discussion}$ and conclude in $\mathrm{Sect.}\,\ref{sec: conclusions}$.

\section{Methodology}
\label{sec: methodology}
In this section, we discuss the code and methodology used to calculate our grid of structure models. The parameter space varied in this work, and other relevant inputs in our structure models are detailed.

\subsection{MESA code}
\label{sec: mesa_code}
We use the MESA structure and evolution code (version r15140) to build our grid of structure models\footnote{\href{https://sourceforge.net/projects/mesa/files/releases/mesa-r15140.zip/download}{https://sourceforge.net/projects/mesa/files/releases/mesa-r15140.zip/download}}. The MESA code solves the five fundamental time-dependent, stellar structure equations with appropriate initial and boundary conditions \citep{Kipp1969} and is capable of simulating the life evolution of a wide range of astrophysical objects including stars, giant planets, brown dwarfs, SNe explosions, to name a few \citep{MESA11, MESA13, MESA15, MESA17, MESA19}. While MESA can evolve a wide range of objects, at its core, it is a structure code. This means that by ensuring the time-dependent terms become negligible, one can obtain a stable structure for a given set of input mass and composition. 

Before describing the strategy to build our grid of synthetic structure models, we first list the input parameters that are fixed in our grid. The core of massive stars is convectively unstable because radiative diffusion alone is unable to transport the enormous amounts of energy produced in the core. Convection in our models is treated using the standard mixing length theory (MLT) from \citet{MLT68} with a fixed convective mixing length parameter of $\alpha_\mathrm{MLT} = 1.5$. The MLT++ routine is switched off, meaning that the temperature gradient predicted by MLT is used as-is, without any artificial reduction in the temperature gradient.

The total initial metal mass fraction $Z$ is an input and a variable in our parameter space. However, the calculation of the initial He and H mass fractions given $Z$ follows a fixed formula:   $Y = Y_{\text{prim}} + (\Delta Y/\Delta Z)\, \times\, Z$, where the primordial He abundance, $Y_{\text{prim}} = 0.24$ and He enrichment factor, $\Delta Y/\Delta Z = 2$ \citep{Audouze1987, Pols1998}, $X = 1-Y-Z$. The individual metal mass fraction spread among the different metals is according to solar-scaled abundances from \citet{GS98}. These are the initial values of $X,\, Y,\, Z$ and the spread among the individual metals. These values will be modified later by entering a custom chemical profile.

For the reaction network, we use the \texttt{basic.net} that includes eight isotopes. The opacities used are from the OPAL Type 2 opacity tables, which take into account any changes in the mass fractions of metals, mainly carbon (C) and oxygen (O). For the equation of state (EOS), we use the tables available in MESA that are mainly based on the OPAL EOS tables \citep{Rogers2002} plus a blend of other EOS tables \citep[for further details, see][]{MESA11}.

\subsection{Constructing structure models}
\label{ref:constructing_structure}

The general procedure to generate our synthetic structure models is as follows:
\begin{enumerate}
    \item Input a given chemical abundance profile. This includes providing the stratification of all the elements in the reaction network used as a function of mass inside the model. In MESA, this is performed using the \texttt{relax\_composition\_filename} option which relaxes the initial chemical profile to the input custom profile. 
    \item We switch off any changes to the abundances due both to nuclear burning and internal mixing by setting \texttt{dxdt\_nuc\_factor} and \texttt{mix\_factor} options to zero. This ensures that no nuclear burning or mixing processes have altered the abundances, and the structure model we calculate corresponds to the input chemical profile.
    \item For this input chemical profile, we let the model relax to thermal balance by checking whether the actual luminosity stratification inside the model equals the nuclear luminosity stratification within a cut-off threshold of $0.1\%$. That is, only when within the entire model, the time-dependent term in the thermal balance equation falls below $0.1\%$ of the actual luminosity do we consider the structure model to be converged for the given chemical profile. We use the hydrostatic option in MESA, meaning the final structure obtained is also in hydrostatic balance.
    \item Once the model has achieved hydrostatic and thermal balance, we obtain the final structure and read out the value of its surface luminosity.
\end{enumerate}

\subsection{Grid parameters}
\label{sec: grid_param}

To build our grid of synthetic structure models consisting of a He core and an H shell, we vary the following four parameters: (1) total mass, $M_\mathrm{tot}$, (2) total metal mass fraction, $Z$, (3) surface value of the H mass fraction, $X_\mathrm{H}$,  and (4) H profile gradient or slope, $s$, with which the surface H depletes inward and ultimately becomes zero in the He-rich core. The last two parameters are sufficient to build the composition file for the \texttt{relax\_composition\_filename} option as we get a unique $X$ profile from the two parameters. The He profile is then just $1-X-Z$. The $Z$ value (and the individual metal spread within) is kept unchanged. Our approach to constructing the synthetic structure model grid closely follows the techniques developed by \citet{Abel2018} and \citet{Farrell2021}.

In reality, the individual metal mass fractions change, especially carbon, nitrogen and oxygen (CNO) abundances during core-H burning where nitrogen (N) increases in favour of C and O, while the sum remains approximately constant. For the models here, we ignore this change and fix the metal mass fractions in the model to their ZAMS values. However, we perform tests in $\mathrm{Appendix}\,\ref{appendix: CNO_testing}$  where the CNO abundances are changed to more realistic CNO-cycle equilibrium values typically obtained in the He cores instead of the initial CNO abundances. The maximum deviation in the predicted luminosities due to the CNO abundance assumption we make is roughly 0.02 dex, with typical values of the order of 0.01 dex, which is negligible.

\begin{figure}
    \includegraphics[width = \columnwidth]{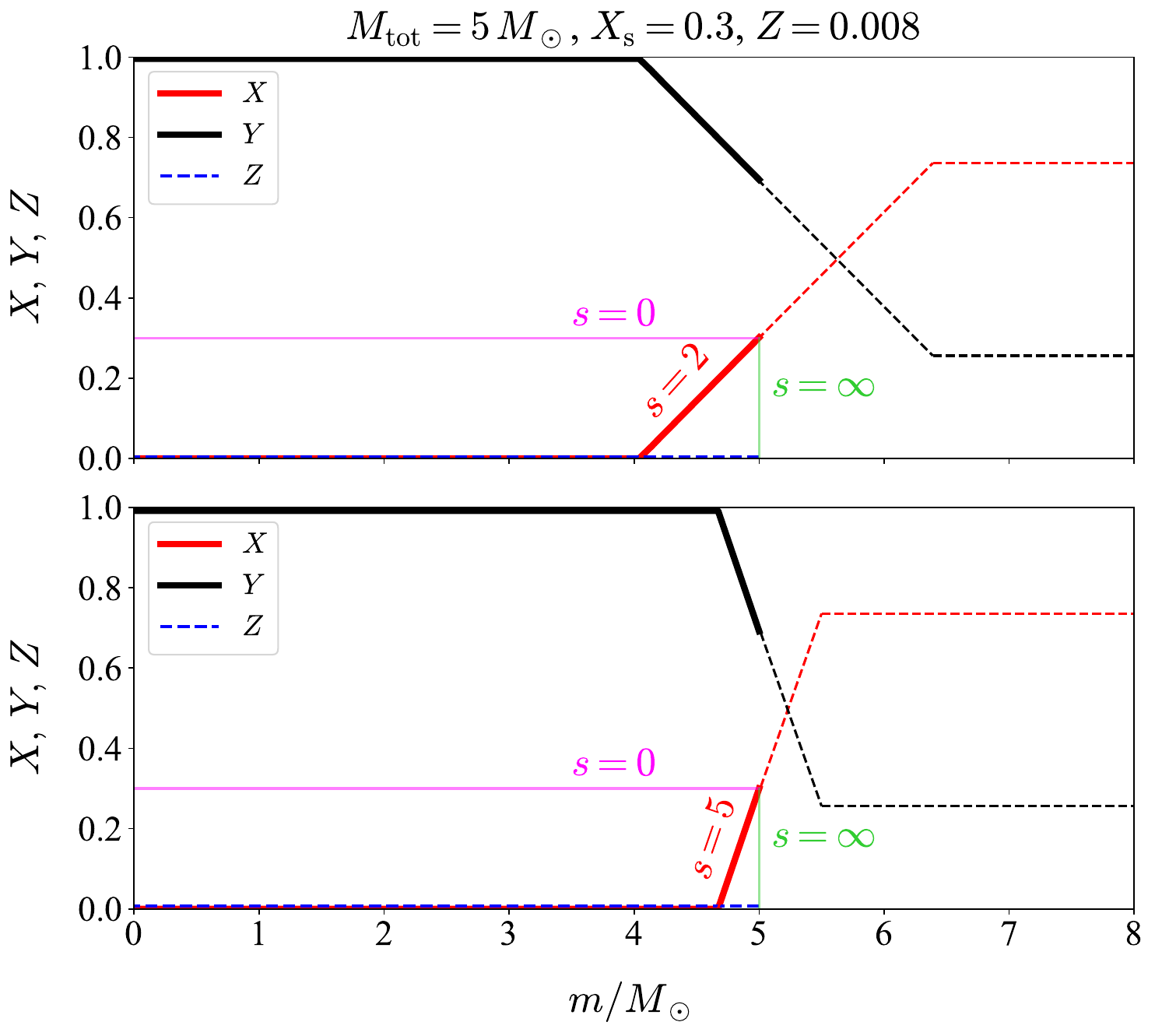}
    \caption{Chemical abundance profile of two synthetic models showcasing different H/He slopes. The mass fractions of hydrogen ($X$), helium ($Y$), and metals ($Z$) are shown in red, black and blue respectively.  The top subplot has a H-profile slope of $s=2$, while the bottom subplot has a higher slope of $s=5$, with all other inputs fixed to the values shown in the title. The two extreme slope values of $s=0$ and $s=\infty$ are also marked in magenta and green respectively. The red and black dashed lines indicates the supposed H and He profile prior to stripping either by wind mass loss or binary stripping.
}
    \label{fig: chem_profile}
\end{figure}

In $\mathrm{Fig.}\,\ref{fig: chem_profile}$, we show the internal profile variation of $X,\,Y$ and $Z$ as a function of stellar mass of two example synthetic models. The two models have a fixed total mass of $M_\mathrm{tot} = 5\,M_\odot$, metal mass fraction $Z$ of $0.008$ and surface H mass fraction of $X_\mathrm{H} = 0.3$. The two models only differ in their H-profile slope values. The definition of the slope $s$ is adopted from \citet{Abel2018, Abel2019}, which is given by
\begin{equation}
\begin{array}{c@{\qquad}c}
s = \dfrac{dX}{dQ}
\end{array}
\label{eq: slope}
\end{equation} 
where $Q$ is a normalised mass-coordinate going from $0$ at the centre to $1$ at the location where the extrapolated H mass fraction value equals the initial value of $X_\mathrm{H}$, which in this case is $1-0.256-0.008 =  0.736$. The He-core mass in the top sub-plot is roughly $4.04\,M_\odot$, and the mass value where the extrapolated H mass fraction equals the initial $X$ is $6.39\,M_\odot$. A quick calculation reveals the slope with which the surface value of $X$ depletes inwards which is $s = (0.736-0)/(1-4.04/6.39) = 2$. Similar calculation can be performed to show that the bottom sub-plot has a H slope of $s=5$.

We now discuss the parameter space of our grid. For structures consisting of a He core and an H shell (i.e., $0 < s < \infty$), the $M_\mathrm{tot}$ values ranges from $1-18\,M_\odot$. Surface $X_\mathrm{H}$ values ranges from 0.01 to 0.7. For the slope $s$, we consider the range between $1$ and $30$. Additional values of $s$ from $0.75$ to $0.9$ are also considered for the $X_\mathrm{H} = 0.01, 0.03,$ and $0.05$ models as their maximum luminosity occurs for slope values $s < 1$. Although such low $s$ values might not be realistic from an evolutionary perspective (see $\mathrm{Fig.}\,\ref{fig: slope_evol}$), we include structure models with $s < 1$ for these specific $X_\mathrm{H}$ values to ensure completeness while providing the maximum MLRs across the full $X_\mathrm{H} = 0.01-0.7$ range.  

Additional structure models were run with slope $s=0$, corresponding to chemically homogeneous models, for which $M_\mathrm{tot}$ ranges from $1-40\,M_\odot$ and $X_\mathrm{H}$ from $0.01-0.7$. Pure-He models with $s=\infty$ were also run, with $M_\mathrm{tot}$ ranging from $1-40\,M_\odot$. For all the models above, two metallicity values were considered, $Z = 0.008$ and $0.004$, which roughly cover the metallicity content of the MCs. A total of $5910$ structure models were run.

\subsection{Values of slope $s$}
\label{sec: slope_values_evo}

The slope of the H profile inside massive stars is affected by various factors, including the receding convective core during the MS, mixing processes such as semiconvection and overshooting that can alter the chemical abundances as well as deep convective regions that form in the envelope during RSG evolution. Multiple studies in the previous literature have investigated the impact of different processes on the slope of the H profile and ultimately the configuration favoured, for example, the red to blue supergiant (BSG) population \citep{Langer1983, Stothers1992, Abel2019, Higgins2020, Farrell2021}. 

\begin{figure*}
    \includegraphics[width = \textwidth]{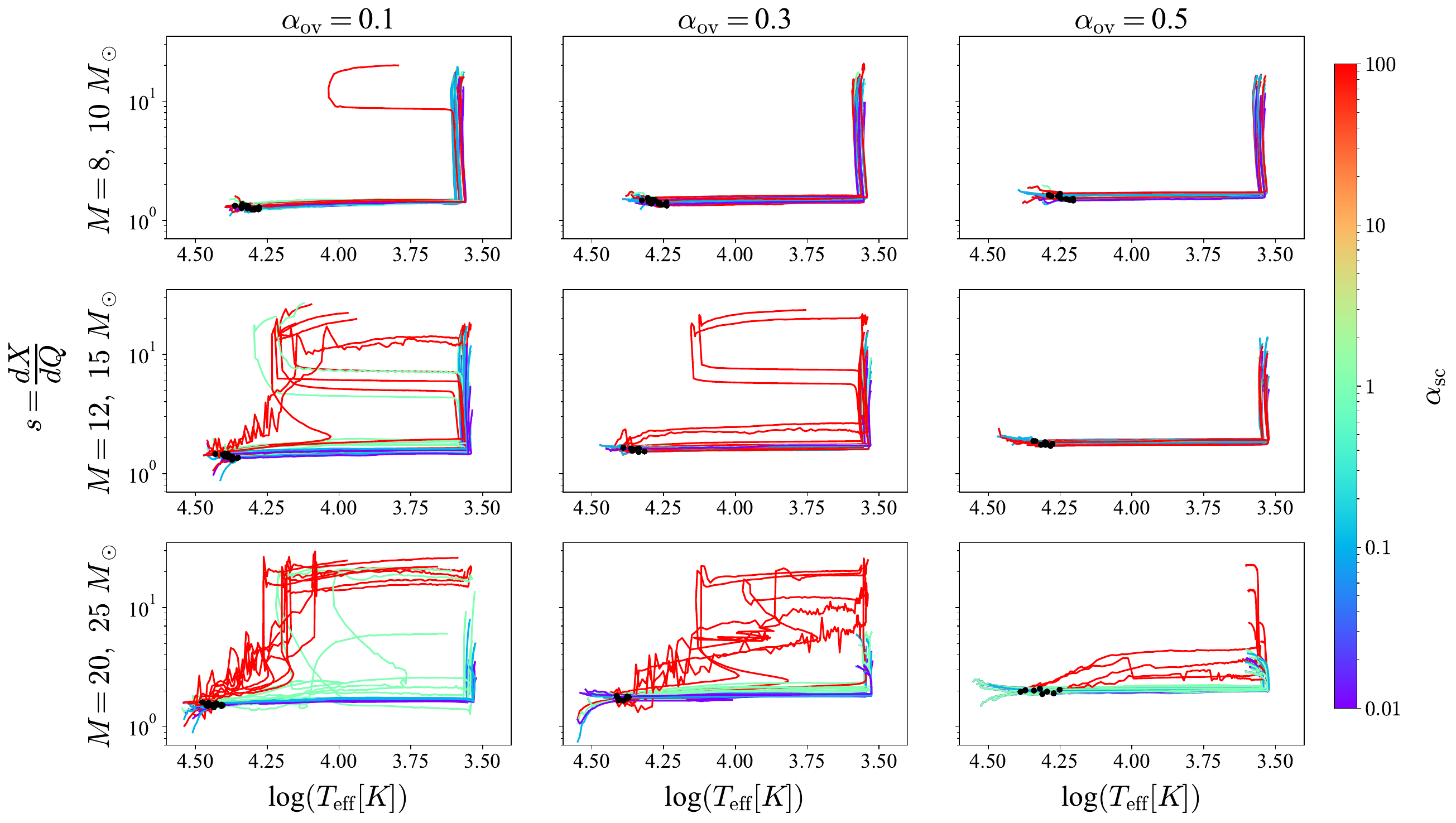}
    \caption{Evolution of H-profile slope $s = dX/dQ$ as a function of effective temperature. The evolutionary model grid is built by varying the initial mass, convective overshooting, semiconvective mixing, rotation and initial metallicity (see text for the ranges spanned by these parameters). The sub-panels are divided based on the input overshooting and initial mass ranges. The tracks are colour-coded based on the input semiconvective efficiency. The black dots mark the end of the MS where central H mass fraction $X_\mathrm{c}$ falls below 0.01.
}
    \label{fig: slope_evol}
\end{figure*}

To obtain a realistic range of possible slope values, we run a dedicated grid of evolutionary models. This grid differs from the structure model grid above, as chemical abundances are allowed to change and evolve by nuclear processes and diffusion. The evolutionary models are run until the end of core-He burning, defined as the point when the central He mass fraction drops below $Y_\mathrm{c} < 0.01$.

The following inputs are varied in our evolution grid. The initial mass ranges from $8$ to $25\,M_\odot$. The convective boundary mixing is in the form of step overshooting where the radius of the overshooting region is defined by a free parameter factor of the local pressure scale height, with the free parameter taking the values $\alpha_\mathrm{ov} = 0.1, 0.3$, and $0.5$. We also vary the efficiency of semiconvective mixing in regions of chemical gradient with diffusive mixing from \citet{Langer1983} where we vary the free parameter coefficient $\alpha_\mathrm{sc} = 0.01, 0.1, 1$ and $100$. Two different rotation rates are tested of $\Omega/\Omega_\mathrm{crit} = 0$ and $0.4$ with rotational-induced instabilities from \citet{Heger2000}. The total initial metallicity values used are $Z = 0.008$ and $0.004$.

The slope $s$ of the H profile is obtained by fitting a straight line through the H mass fraction profile. Specifically, we consider the H mass fraction increasing upwards from the edge of the He core (for models beyond the MS, $X=0$ at the edge) up to the minimum value between the surface H mass fraction and $0.4$, and fit a straight line to this region. In these detailed evolutionary models, small and large step-like features can develop in the H profile due to semiconvective and fully convective regions that transport H from the overlying envelope. Locally, the slope within these steps can be very steep, but on average across the increasing H profile, it lies within the range $s\approx1$ to $s\approx30$. While a linear fit may not capture the local slope variations across these steps, the purpose of running a dedicated evolutionary grid is to estimate an approximate slope range as input in our synthetic structure grid. Ultimately, the effects of such step-like features on the predicted minimum and maximum MLRs are minimal. For instance, more H-rich material in the steps tends to reduce the luminosity for a given mass, resulting in luminosities that fall between our predicted minimum and maximum, with the two extremes largely unaffected.

In $\mathrm{Fig.}\,\ref{fig: slope_evol}$, we plot the best-fit slope $s$ as a function of surface effective temperature. Since temperature typically decreases during the MS as well as during the so-called HG expansion phase where stars eventually become BSGs or RSGs, we can consider the temperature in the abscissa as a proxy for age. We observe a common trend in our models across all overshooting values and mass ranges. Up until the end of the MS, the models exhibit a shallow slope of $s \approx 1$–$3$, with slope values close to unity for the lowest overshooting and the lowest mass range considered here. Beyond the MS, H burns in a shell around the He core, gradually moving outward through the H profile, thereby increasing the slope. Steeper slopes of $s \approx 3-30$ are seen beyond the MS, either already during the HG phase or later during core-He burning. 

During and towards the end of the MS, a small but systematic increase in the slope is seen with increasing MS overshooting, consistent with the findings of \citet{Abel2018} (see their Table 3). The initial mass also has a weak influence on the slope value at the end of the MS, with slightly higher slopes at higher masses due to the larger convective cores during the MS. Beyond the MS, however, the supergiant structure during core-He burning -- and consequently the slope evolution -- differs significantly across models with varying MS overshooting and initial mass, depending on the efficiency of semiconvective mixing.

Models with low semiconvective mixing stabilise as RSGs ($\log(T_\mathrm{eff}/\mathrm{K}) \sim 3.6$) beyond the MS, regardless of overshooting or initial mass. Their H-profile slope gradually increases during core-He burning in the RSG phase. Conversely, in models with very efficient semiconvective mixing, steeper slopes are already observed at higher temperatures, between $\log(T_\mathrm{eff}/\mathrm{K}) \sim 4.3-4.0$, where the stars stabilise as BSGs. 

However, not all high semiconvective models stabilise as BSGs. For high overshooting values, the semiconvective region above the core is suppressed, and the star rapidly crosses the HG regardless of semiconvective efficiency. These models then stabilise as RSGs, with most of the slope increase occurring during the RSG phase. In contrast, for low overshooting, high semiconvective models begin core-He burning already as BSGs, resulting in an early slope increase at $\log(T_\mathrm{eff}/\mathrm{K}) \sim 4.3-4.0$. A similar trend is seen with increasing initial mass: high semiconvective models tend to stabilise as BSGs and initiate core-He burning in that phase. In comparison, the rotation and metallicity ranges tested here have a minimal effect on the slope evolution beyond the MS. Overall, evolutionary models predict typical slope values in the range $s \approx 1-30$.

\subsection{Hydrodynamically consistent $\texttt{PoWR}^\textsc{hd}$ atmosphere models}

Given the evolutionary significance of partially-stripped stars, it is important to understand their wind structure and mass-loss properties. In the next section, we will demonstrate that structures resulting from partial stripping can have higher luminosities than their fully-stripped, pure-He counterparts of the same mass. This could mean a stronger wind due to the higher Eddington parameter, which could impact their evolution. In this work, we explore the wind structure of partially-stripped configurations by computing a small sample of hydrodynamically consistent atmosphere models using the non-LTE code \texttt{PoWR} \citep{Grafener2002, HG2003, Sander2015}.

The standard version of the $\texttt{PoWR}$ code models a spherically symmetric, expanding, non-grey atmosphere with a stationary wind outflow, where the velocity field and the mass-loss rate are provided as inputs. The code then iteratively solves the radiative transfer in CMF \citep[based on the concepts of][]{Mihalas1975, Mihalas1976}, and a set of statistical equilibrium equations, fully coupled in spatial and frequency domain, until a self-consistent atmosphere model is obtained for the prescribed velocity field and input mass-loss rate. 

The hydro-version of the code removes the wind stratification inputs by solving the hydrodynamical equation of motion. These $\texttt{PoWR}^\textsc{hd}$ models can thus calculate the velocity field consistent with the different forces in the atmosphere, while mass loss is typically fixed implicitly by conserving the Rosseland continuum optical depth. 

Further details on the setup and usage of the $\texttt{PoWR}$ code and the hydro-version can be found in \citet{Grafener2002, HG2004, Sander2017, Sander2020a} and \citet{Sander2023}. For more details regarding the model inputs for $\texttt{PoWR}^\textsc{hd}$ models presented in this work, see \citet{Sabhahit_VMS2025}.

\section{Mass-luminosity relations}
\label{sec: MLR}

In this section, we provide fit formulae to predict the minimum, maximum and pure-He luminosities for a given total mass and surface $X_\mathrm{H}$, and the inverse problem, and get the minimum, maximum and pure-He masses for a given luminosity and surface $X_\mathrm{H}$. 

\begin{figure}
    \includegraphics[width = \columnwidth]{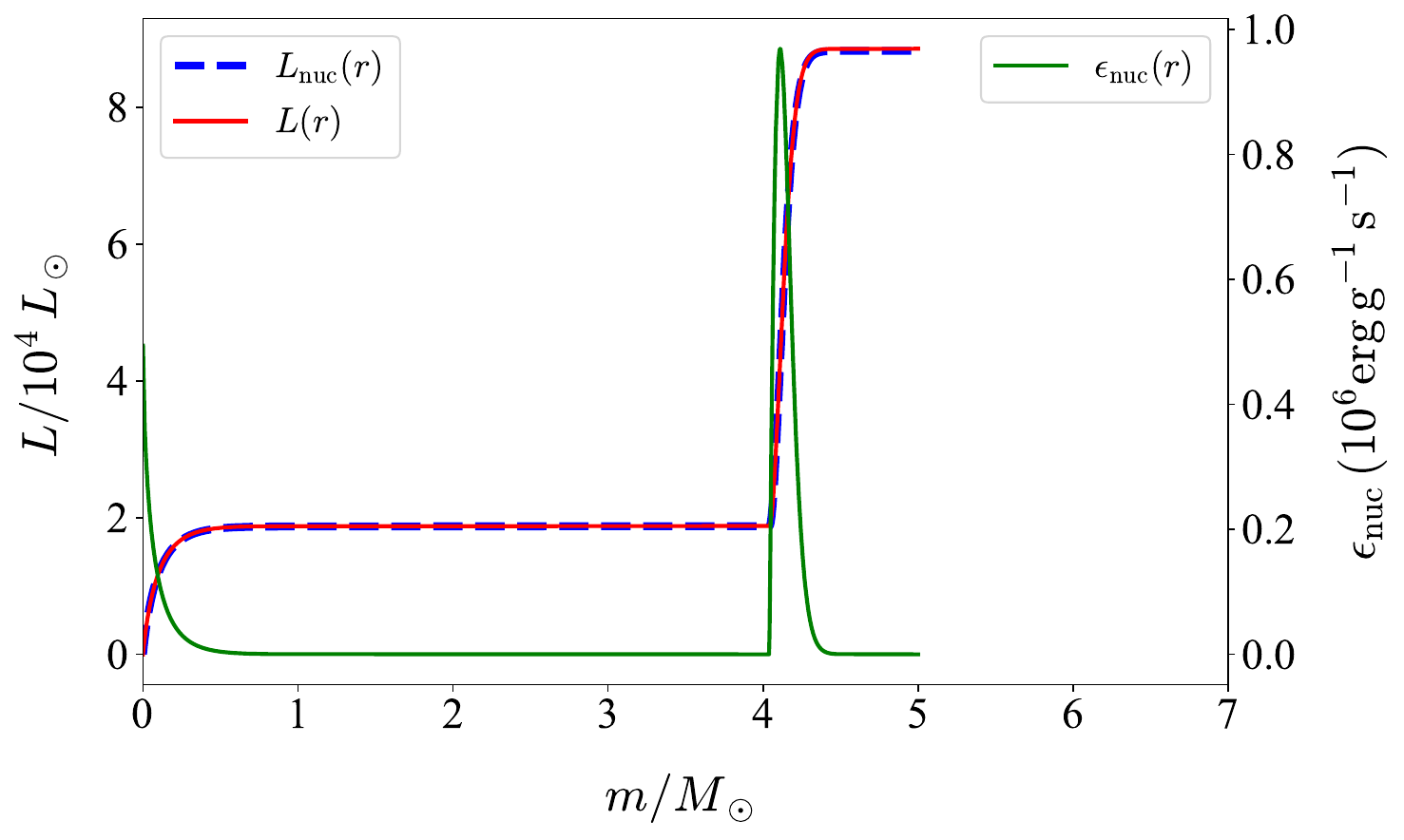}
    \caption{Actual luminosity stratification compared to nuclear luminosity inside a synthetic partially-stripped structure model in thermal balance. The specific nuclear energy generation rate, $\epsilon_\mathrm{nuc}$, in erg/g/s is also shown. This is the same model as in the top sub-plot of $\mathrm{Fig.}\,\ref{fig: chem_profile}$.
}
    \label{fig: chem_profile_l}
\end{figure}

\begin{figure}
    \includegraphics[width = \columnwidth]{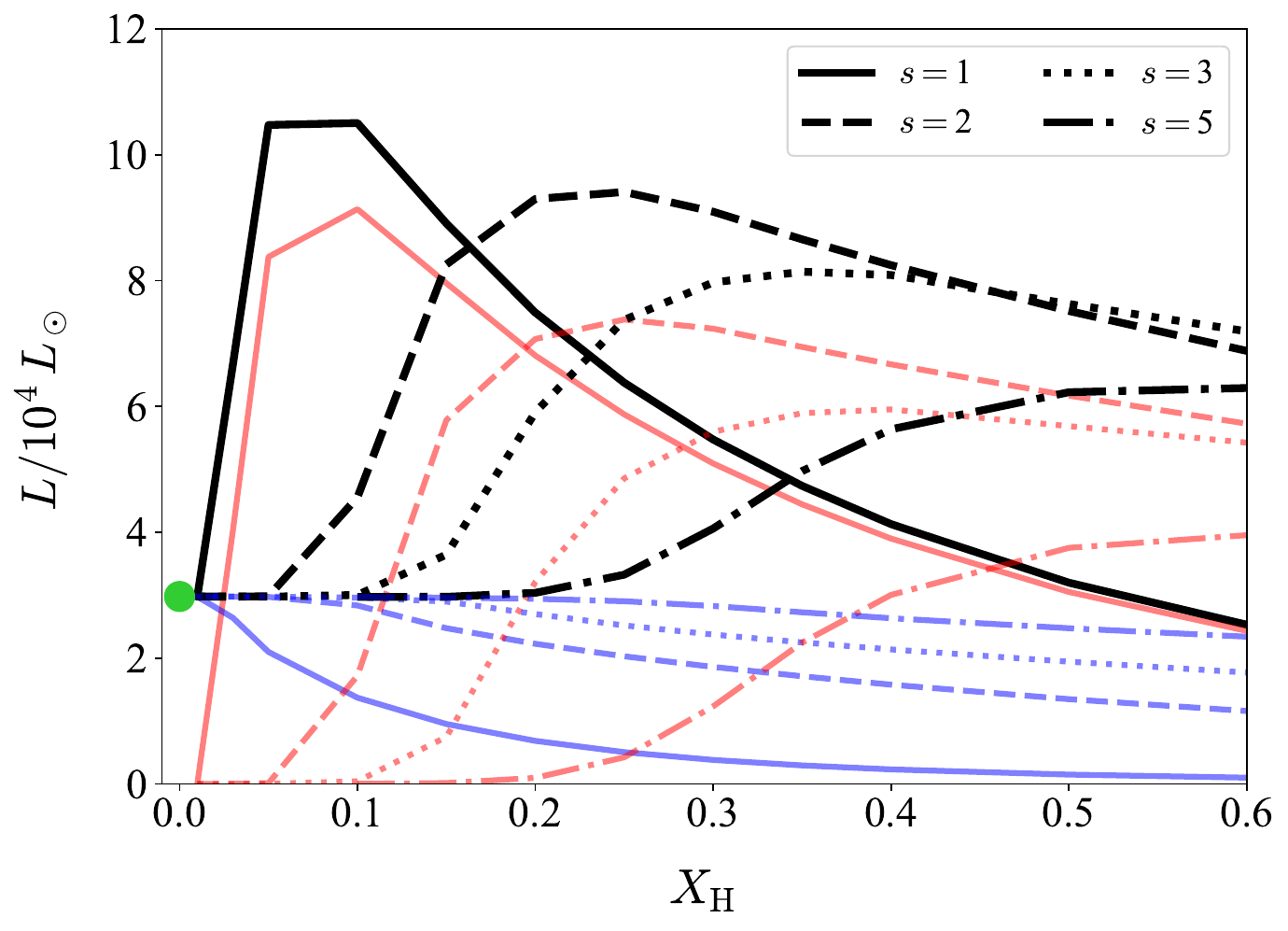}
    \caption{Variation of surface luminosity (black) as a function of $X_\mathrm{H}$ for different values of slope $s$ for $Z=0.008$. The parameter $M_\mathrm{tot}$ is fixed to $5\,M_\odot$. The individual contributions from the He core (blue) and H shell (red) are also shown. The green dot indicates the luminosity of the $5\,M_\odot$ pure-He  model.
}
    \label{fig: H_He_lum_contributions}
\end{figure}

\subsection{Structure of partially-stripped configurations}

\begin{figure*}
    \includegraphics[width = \textwidth]{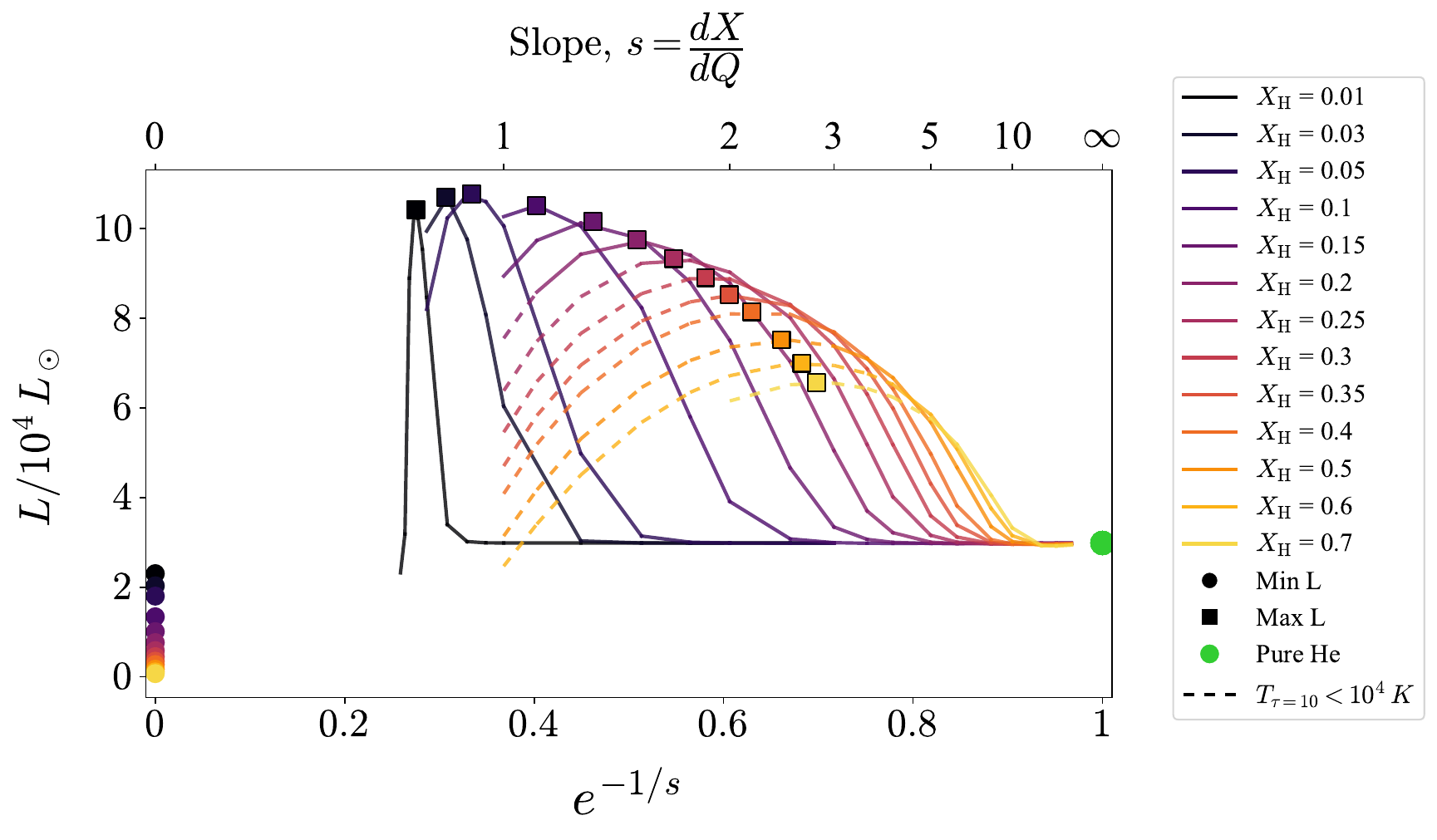}
    \caption{Variation of luminosity as a function of slope $s$ for different values of $X_\mathrm{H}$ for $Z=0.008$. The parameter $M_\mathrm{tot}$ is fixed to $5\,M_\odot$. The abscissa goes from a fully chemically homogeneous model ($s = 0$) to a pure-He model ($s = \infty$). The minimum and maximum luminosities for a given total mass and surface $X_\mathrm{H}$ are marked by coloured dots and squares respectively. The green dot indicates the luminosity of the $5\,M_\odot$ pure-He  model. }
    \label{fig: logL_vs_slope}
\end{figure*}

As already mentioned in the introduction, the reason for revisiting the MLRs for stripped stars lies in the non-intuitive and disproportionate luminosity contribution from the H-burning shell in a typical partially-stripped structure. 

\citet{Graf2011} has provided fit formulae to predict the minimum and maximum luminosities for a given mass and surface $X_\mathrm{H}$ (as well as the inverse problem). As $L \sim  \mu^4 M^3$ from simple homology relations, the minimum luminosity arises in a fully chemically-homogeneous model, where the chemical profile is uniform throughout the star (i.e., $X(m) = X_\mathrm{H}$). Conversely, the maximum luminosity occurs for the pure-He case (i.e., $X(m) = 0$).  

In $\mathrm{Fig.}\,\ref{fig: chem_profile_l}$, we show the actual and nuclear luminosity stratification of a typical partially-stripped structure model in thermal balance. At the surface, the luminosity of the model is $8.858 \times 10^4\,L_\odot$.  The nuclear burning sources are the He-burning core and the H-burning shell, as seen by the two spikes in the specific nuclear energy generation rate ($\epsilon_\mathrm{nuc}$) at $m = 0\,M_\odot$ and $m \approx 4.04\,M_\odot$. The disproportionate contributions from the two sources are evident in the figure. The He core, which occupies approximately $4.04\,M_\odot$, contributes only about one-fourth of the total luminosity. In contrast, the H shell, weighing just $0.96\,M_\odot$, dominates the luminosity by contributing nearly three-fourths of the total luminosity budget. 

In comparison, a pure-He star of the same total mass of $5\,M_\odot$ has a luminosity of $2.978 \times 10^4\,L_\odot$. Clearly the above structure with $s$ somewhere in between $0$ and $\infty$ has more than a factor of two higher luminosity compared to the $s=\infty$ pure-He case for the same total mass. 

The aim below is to predict the maximum luminosity achievable by typical partially-stripped structures for a given total mass and surface $X_\mathrm{H}$. Or to reverse the question, and predict the minimum possible mass for a given luminosity and surface $X_\mathrm{H}$.

\subsection{Total luminosity as a function of surface $X_\mathrm{H}$ and slope}
\label{sec: L_vs_slope}

\gnstext{We begin by showing the non-trivial behaviour of the luminosity with respect to the slope and surface $X_\mathrm{H}$. Let us first consider a model sequence with a fixed $M_\mathrm{tot}$ and $s$, while varying surface $X_\mathrm{H}$. In $\mathrm{Fig.}\,\ref{fig: H_He_lum_contributions}$, we show the total surface luminosity as a function of surface $X_\mathrm{H}$ for four different values of $s$. The total mass is fixed at $5\,M_\odot$. The individual contributions from the He core and H shell to the total luminosity are also shown. The green dot at $X_\mathrm{H} = 0$ marks the aforementioned pure-He luminosity of $2.978 \times 10^4\,L_\odot$.}

\gnstext{We notice that the variation of the total luminosity as a function of $X_\mathrm{H}$ is non-monotonic, first increasing and then decreasing. There is a threshold surface $X_\mathrm{H}$ beyond which the H shell begins to dominate the total luminosity. For example, for $s = 2$, the H-shell luminosity is practically zero for $X_\mathrm{H} \lesssim 0.05$ as the H content is too low for any meaningful contribution to the total nuclear burning output. Above $X_\mathrm{H} \sim 0.05$, the H-shell luminosity rapidly grows, and for $X_\mathrm{H} \gtrsim 0.1$, the H shell dominates the total luminosity. Upon reaching a maximum at $X_\mathrm{H} \sim 0.25$, both the H shell and the total luminosity decrease. The decrease in luminosity is due to a combination of increasing opacity and a decreasing mean molecular weight with increasing H content in the envelope} \citep{Farrell2021}. The He-core luminosity, on the other hand, monotonically decreases with increasing $X_\mathrm{H}$. This is because increasing $X_\mathrm{H}$ in a model sequence while fixing both $M_\mathrm{tot}$ and $s$ results in a decreasing core-to-total mass ratio, thereby reducing the luminosity contribution from the He core.

\gnstext{We can now investigate the same model sequence  while fixing $M_\mathrm{tot}$ and surface $X_\mathrm{H}$ and varying the slope instead. In $\mathrm{Fig.}\,\ref{fig: logL_vs_slope}$ we show the variation of luminosity as a function of slope for different values of $X_\mathrm{H}$. The total mass is again fixed at $M_\mathrm{tot} = 5\,M_\odot$. Similar to \citet{Graf2011}, the minimum luminosity for a given surface $X_\mathrm{H}$ occurs at $s = 0$, corresponding to a fully chemically-homogeneous star. From an evolutionary perspective, chemical homogeneous structure can be achieved, for example, due to rapid rotation \citep{Yoon2005, Woosley2006}, but see \citet{Vink_Harries_2017}. }

\gnstext{At the other extreme is the $s = \infty$ pure-He configuration, corresponding to full stripping where the entire H envelope is lost. However, as evident in the figure, the pure-He case does not have the maximum luminosity among all the $5\,M_\odot$ structure models. In fact, we notice the luminosity to first increase with increasing slope, reaches a maximum for a certain value of $s$ between the $s=0$ and $s=\infty$ extremes, and then reduces to asymptotically approach the pure-He luminosity as $s\rightarrow \infty$. }

\gnstext{We can understand this behaviour by starting from the $s = \infty$ case and moving in the reverse direction, decreasing the slope. For a fixed $M_\mathrm{tot}$ and surface $X_\mathrm{H}$, decreasing $s$ means increasing total H content (cf. $\mathrm{Fig.}\,\ref{fig: chem_profile}$). The $s=\infty$ has no H, and as $s$ decreases, the H content increases with the $s=0$ case having the maximum possible H content. Initially as $s$ decreases, the H-burning shell luminosity rapidly increases and begins to dominate. The total luminosity increases and reaches a maximum. Beyond this maximum, the competing effects of higher opacity and lower mean molecular weight} due to the increasing H content dominates and the total luminosity decreases. 

The maximum luminosity achievable from a partially-stripped structure can be roughly $3-4$ times the luminosity of the fully-stripped pure-He case. A similar behaviour in the variation of luminosity with slope $s$ can also be observed across the entire $M_\mathrm{tot}$ range tested in this work (see $\mathrm{Appendix}\,\ref{appendix: diff_mass}$ for similar plots at different $M_\mathrm{tot}$).  It is clear from these model sequences that the luminosity is a non-trivial function of surface $X_\mathrm{H}$ and slope. 

An important note on surface temperatures is warranted here. The dashed lines in $\mathrm{Fig.\,\ref{fig: logL_vs_slope}}$ indicate models with effective temperatures at $\tau_\mathrm{Ross} = 10$ below $10\,\mathrm{kK}$. The maximum luminosity at sufficiently high surface $X_\mathrm{H}$ occurs at temperatures cooler than $10\,\mathrm{kK}$. While this may introduce potential discrepancies when comparing these structure models with observed partially stripped stars -- whose temperatures are typically hotter \citep[see, e.g.,][]{Ramachandran2023, Ramachandran2024} -- it is important to note that the both the measurements and the predictions of the surface temperatures are subject to uncertainties. Depending on the binary configurations (e.g., flux ratios, spectral types) and available wavelength ranges, the temperature diagnostics for a quantitative spectroscopic analysis of the (partially) stripped star can be limited. On the other hand, 1D structure models of partially-stripped stars show inflated morphologies owing to their high luminosity-to-mass and low gas-to-total pressure ratios, despite having only a few solar masses \citep{Sabhahit2025}. Consequently, the predicted radius (and hence the effective temperature) can be highly uncertain and is strongly influenced by the model’s proximity to the Eddington limit and the treatment of convection in these layers. Whether such inflated layers occur in nature remains an open question. Given the high luminosity-to-mass ratios characteristic of partially stripped configurations which could lead to inflation, and the inherent uncertainties associated with temperature estimates, we do not assign the same diagnostic significance to temperature as we do to luminosity and mass in our model-observation comparisons in $\mathrm{Sect.\,\ref{sec: discussion}}$. Accordingly, our analysis is primarily based on luminosity and mass.

\subsection{Minimum, maximum and pure-He luminosities for given mass and hydrogen abundance}

\begin{figure*}[h!]
    \includegraphics[width = \textwidth]{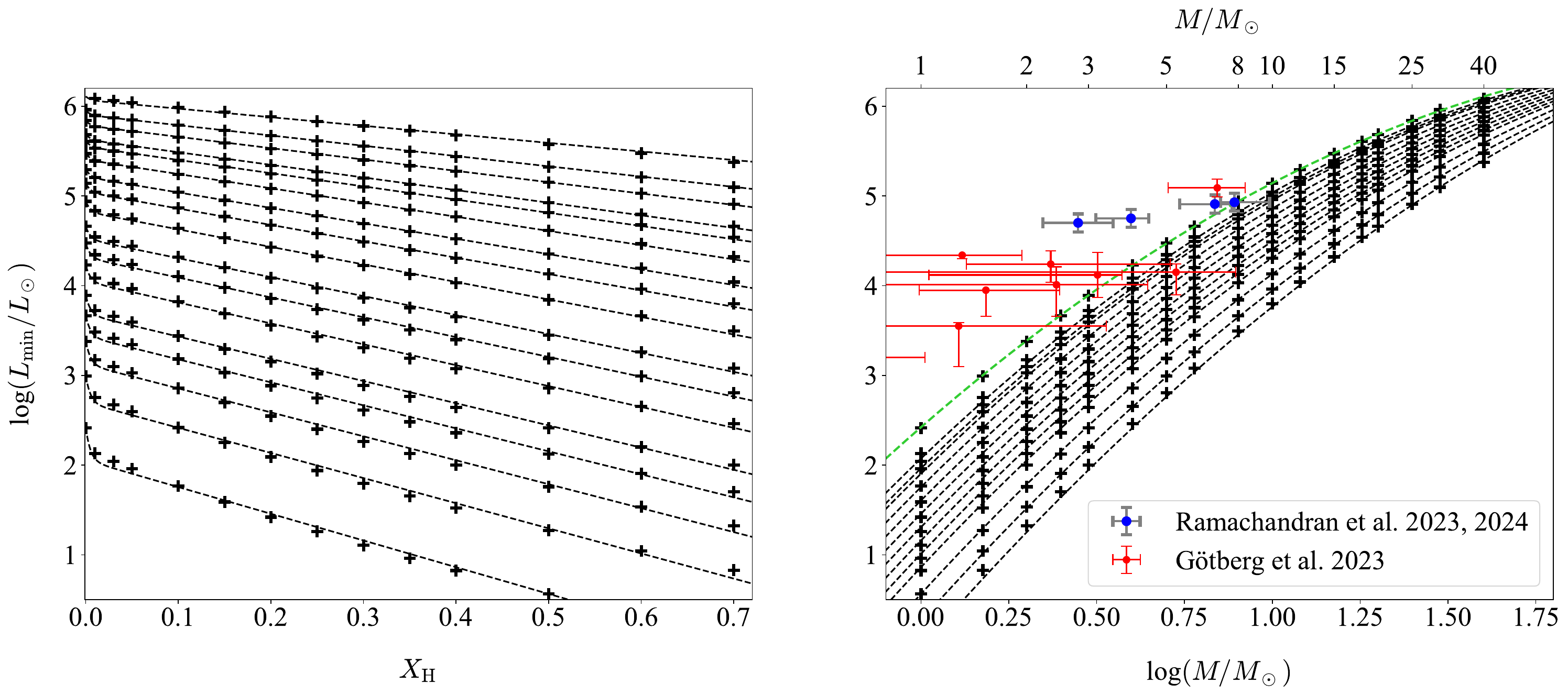}
    \caption{Minimum luminosity for a given total mass and surface $X_\mathrm{H}$. \textit{Left:} Minimum luminosity plotted against surface $X_\mathrm{H}$ for total masses ranging from $1\,M_\odot$ to $40\,M_\odot$.  \textit{Right:} Minimum luminosity plotted against total mass for $X_\mathrm{H}$ values ranging from 0 to 0.7. The plus signs indicate minimum luminosities calculated from fully chemically homogeneous structure models, while the dashed black lines represent our best-fit relations from Eq. \ref{eq: L_giv_M_X}. The green dashed line in the right sub-panel corresponds to the MLR for pure-He models. Over-plotted are stripped star luminosities and spectroscopic masses from \citet{Gotberg2023} and \citet{Ramachandran2023, Ramachandran2024}.
}
    \label{fig: Min_L_given_M_X}
\end{figure*}

By constructing similar luminosity versus slope plots for different values of $M_\mathrm{tot}$ in our grid, we can predict the minimum, maximum and pure-He luminosities as a function of surface $X_\mathrm{H}$ and $M_\mathrm{tot}$. In $\mathrm{Fig.}\,\ref{fig: Min_L_given_M_X}$, we show the minimum luminosity, where the plus symbols are our model estimates for the minimum luminosity, while the black dashed lines are obtained by our best-fit formula:
\begin{equation}
\begin{split}
\log\left(L/L_\odot\right) & = \,\,  F_1 + F_2 \log\left(M/M_\odot\right) + F_3 \log\left(M/M_\odot\right)^2 \\[8pt] 
& + \Bigg[F_4 + F_5 \log\left(M/M_\odot\right) + F_6 \log\left(M/M_\odot\right)^2 \Bigg] \;X_\mathrm{H} \\[8pt] 
& + \Bigg[F_7 + F_8 \log\left(M/M_\odot\right) + F_9 \log\left(M/M_\odot\right)^2 \Bigg]\; 
e^{-X_\mathrm{H}/F_{10}}
\end{split}
\label{eq: L_giv_M_X}
\end{equation}
where the best-fit coefficients for $F_\mathrm{1}$ through $F_\mathrm{10}$ are listed in $\mathrm{Table}\,\ref{tab:parameters}$. The pure-He case is obtained by setting $X_\mathrm{H} = 0$ in the above relation. The validity ranges of this formula to predict the minimum and pure-He luminosities are: $1 \leq M_\mathrm{tot}/M_\odot \leq 40$, and $0 \leq X_\mathrm{H} \leq 0.7$. 

\gnstext{In $\mathrm{Fig.}\,\ref{fig: Min_L_given_M_X}$, we also over-plot the luminosity and spectroscopic masses of observed stripped-envelope stars in the MCs from \citet{Gotberg2023} and \citet{Ramachandran2023, Ramachandran2024} respectively}\footnote{The observations include objects from both the Large and Small MCs. While the grid is for the $Z=0.008$. The comparison is valid as changing $Z$ to $0.004$ changes the minimum, maximum and pure-He luminosities by roughly $ \sim0.01$ dex.}. \gnstext{Most of these partially-stripped objects have non-negligible H abundance at their surface but still lie above the pure-He line, that is, these objects are clearly \textit{over-luminous} compared to the pure-He configuration for their mass. }

\begin{figure*}[h!]
    \includegraphics[width = \textwidth]{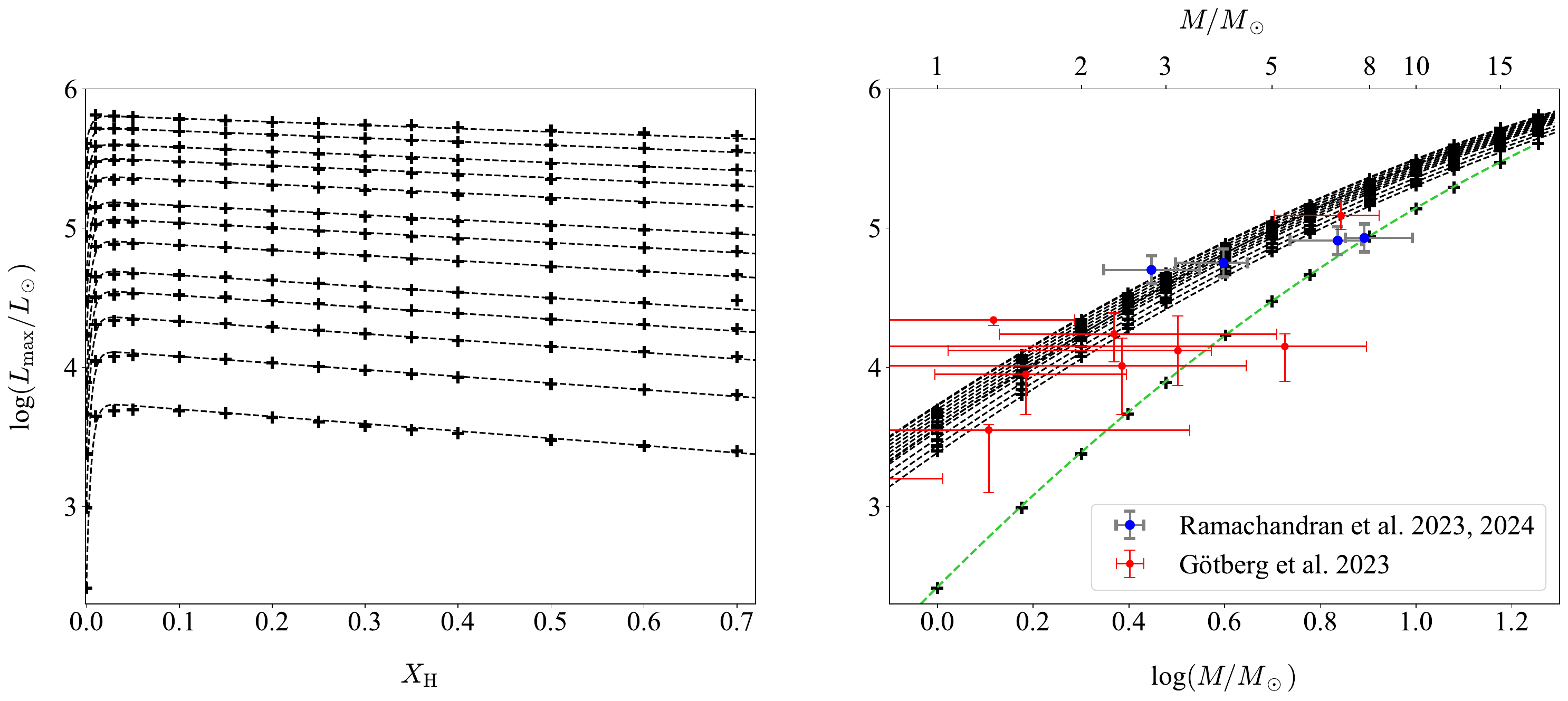}
    \caption{Maximum luminosity for a given total mass and surface $X_\mathrm{H}$. \textit{Left:} Maximum luminosity plotted against surface $X_\mathrm{H}$ for total masses ranging from $1\,M_\odot$ to $18\,M_\odot$.   \textit{Right:} Maximum luminosity plotted against total mass for $X_\mathrm{H}$ values ranging from 0 to 0.7. The plus signs indicate maximum luminosities achievable from partially-stripped structure models, while the dashed black lines represent our best-fit relations from Eq. \ref{eq: L_giv_M_X}. The green dashed line in the right sub-panel corresponds to the MLR for pure-He models. Over-plotted are stripped star luminosities and spectroscopic masses from \citet{Gotberg2023} and \citet{Ramachandran2023, Ramachandran2024}.
    }
    \label{fig: Max_L_given_M_X}
\end{figure*}

The functional form of our fitting formulae is very similar to that provided in \citet{Graf2011}, which involved linear terms in $X_\mathrm{H}$ and quadratic terms in $M_\mathrm{tot}$. We observe in the left panels of $\mathrm{Figs.}\,\ref{fig: Min_L_given_M_X}$ and $\ref{fig: Max_L_given_M_X}$ that the pure-He case does not strictly follow the linear trend of minimum and maximum luminosities with surface $X_\mathrm{H}$. In \citet{Graf2011}, two separate relations were provided: one for non-zero $X_\mathrm{H}$ chemically homogeneous case and another for the pure-He case. To ensure that a single formula which applies for non-zero values of $X_\mathrm{H}$, can also be used for the pure-He case, we use an additional exponential term for $X_\mathrm{H}$ which takes care of the non-linear behaviour below $X_\mathrm{H} \approx 0.01$. We fix the coefficient $F_{10}$ to 0.005, such that for $X_\mathrm{H}$ values on the order of 0.01 and below, the exponential term captures the deviation from linearity.

In $\mathrm{Fig.}\,\ref{fig: Max_L_given_M_X}$, we show the maximum luminosity as a function of surface $X_\mathrm{H}$ and $M_\mathrm{tot}$. To predict the maximum luminosity, we use the same functional form as $\mathrm{Eq.}\,\ref{eq: L_giv_M_X}$, but with different best-fit coefficients which are listed in $\mathrm{Table}\,\ref{tab:parameters}$. The validity ranges of this fit formula to predict the maximum luminosity are: $1 \leq M_\mathrm{tot}/M_\odot \leq 18$ and $0 \leq X_\mathrm{H} \leq 0.7$. 

\gnstext{We can compare our maximum luminosity curves to the observed partially-stripped luminosities. If these curves indeed represent the maximum possible luminosity for a given mass, the observed luminosities should lie below the maximum we predict. And within the error bars, the objects correctly fall below their maximum luminosity curves.}

The same exercise is repeated for $Z = 0.004$ and the best-fit coefficients are listed in $\mathrm{Table}\,\ref{tab:parameters}$. The maximum fitting error from using the above fit formula is about 0.03 dex (0.05 dex) for $L_\mathrm{max}$ ($L_\mathrm{min}$) with typical errors of the order 0.01 dex. The corresponding diagrams for $Z = 0.004$ are shown in $\mathrm{Appendix}\,\ref{appendix: SMC_Z}$.

The relation in $\mathrm{Eq.}\,\ref{eq: L_giv_M_X}$ predicts the minimum, maximum and pure-He luminosities for a given total mass and surface $X_\mathrm{H}$. Predicting the maximum luminosity from this relation, however, does not require the actual value of the slope $s$ for which the maximum luminosity occurs. But knowing the value of $s$ which maximises the luminosity could be useful for reconstructing structure models in the future. So we provide best-fit relations to predict this value of slope in $\mathrm{Appendix}\,\ref{appendix: slope_fits}$. 

In order to get the minimum (or maximum) mass for a specified luminosity and surface $X_\mathrm{H}$, the maximum (or minimum) luminosity formula needs to be inverted\footnote{A python script and an online calculator to calculate minimum, maximum and pure-He mass-luminosity relations are available at \href{https://mdot-com.github.io/scripts/}{https://mdot-com.github.io/scripts/}.}.  The same equation in $\mathrm{Appendix}\,\ref{appendix: slope_fits}$ to find the value of $s$ which maximises the luminosity can be employed to find the value of $s$ which minimises the mass for a given luminosity.

\subsection{Hypothetical example case}

In this section, we work through an example case using our MLRs and compare them with estimates from previous work of \citet{Graf2011}. First, we demonstrate the luminosity predictions from our relations using a fiducial example with  $M_\mathrm{tot} = 10\,M_\odot$, $X_\mathrm{H} = 0.1$ and $Z = 0.008$. Three separate luminosities can be obtained, corresponding to the minimum ($s = 0$), the pure-He case ($s = \infty$) and the maximum ($s = 1.18$), which the formula predicts to be log$(L/L_\odot) =  4.86, \,5.144$, and $5.478$ respectively. In the upper section of $\mathrm{Table}\,\ref{tab: compare_works}$, we compare these absolute values of luminosity to those predicted by \citet{Graf2011} (using their relation numbers 2 and 7, see their Table A.1). The minimum luminosity predicted by the two works differs by less than 0.05 dex. We can also compare the pure-He case which agrees within 0.01 dex. However, there is a difference of approximately 0.34 dex in the predicted maximum luminosity which is roughly a factor of two compared to the pure-He case. 

Now, to invert the problem, we consider a luminosity of $\mathrm{log}(L/L_\odot) = 5.144$ and $X_\mathrm{H} = 0.1$. Using the python script, we obtain three different masses, the maximum mass ($s = 0$), the pure-He mass ($s = \infty$) and the minimum mass ($s = 1.15$) with values 13.428, 9.995 and 5.838 $M_\odot$ respectively. These masses are compared with the relations from \citet{Graf2011} (using their equations 12 and 17 from their Table A.1) in the lower section of $\mathrm{Table}\,\ref{tab: compare_works}$.

As a quick sanity check of the inversion script, we note that the chosen luminosity of $\mathrm{log}(L/L_\odot) = 5.144$ corresponds to a pure-He model with a mass of $10\,M_\odot$. Thus, when inverting the problem, the script should predict a pure-He mass very close to $10\,M_\odot$, which is indeed the case.

As for the absolute values of the masses, the maximum mass shows excellent agreement, differing by about $0.4\,M_\odot$. The pure-He mass predictions also show excellent agreement. The minimum mass differs by roughly a factor of two when compared to the pure-He case. The minimum mass occurs for a slope value of $s = 1.15$. Such low values of the slope that maximise the luminosity (or minimize the mass) will have implications for observed stripped stars, which will be discussed in $\mathrm{Sect.}\,\ref{sec: discussion}$.

Finally, if instead we use the $Z = 0.004$ fit coefficients for the same set of inputs, we get minimum, pure-He and maximum luminosities of log$(L/L_\odot) = 4.872,\, 5.144$ and $5.477$ respectively. These values are roughly 0.01 dex within the values predicted for $Z= 0.008$ case. For the inverse problem, the three masses we obtain at $Z=0.004$ are 13.296, 9.996, and $5.927\,M_\odot$, which are within an order of $0.1\,M_\odot$ of the $Z=0.008$ values. For $Z$ other than 0.008 and 0.004, an interpolation or extrapolation can be performed. 

\begin{table}[b!]
\centering
\footnotesize
\caption{Comparison of luminosity and mass predictions from our work and \citet{Graf2011}. }
\begin{tabular}{|p{3.2cm}|p{1.4cm}|p{2.9cm}|}
\hline
\multicolumn{3}{|c|}{\textbf{$M_\mathrm{tot}=10\,M_\odot, X_\mathrm{H} = 0.1$}} \\ \hline
 & {This work} & {\citet{Graf2011}} \\ \hline
log$(L_\mathrm{min}/L_\odot)$ at $s = 0$  & 4.86 & 4.819  \\
log$(L_\mathrm{He}/L_\odot)$ at $s = \infty$ &  5.144 &  5.133  \\
log$(L_\mathrm{max}/L_\odot)$ at $s = 1.18$ & 5.478 & $-$  \\
\hline
\end{tabular}

\vspace{0.5cm}

\begin{tabular}{|p{3.2cm}|p{1.4cm}|p{2.9cm}|}
\hline
\multicolumn{3}{|c|}{{log$(L/L_\odot)=5.144, X_\mathrm{H} = 0.1$}} \\ \hline
 & {This work} & {\citet{Graf2011}} \\ \hline
$M_\mathrm{max}$ at $s=0$  & 13.428 & 13.829  \\
$M_\mathrm{He}$ at $s = \infty$ &  9.995 &  10.109 \\
$M_\mathrm{min}$ at $s = 1.15$ & 5.838 & $-$  \\
\hline
\end{tabular}
\tablefoot{\textit{Top:} Minimum, maximum and pure-He luminosity prediction for a fixed mass of $M_\mathrm{tot} = 10\,M_\odot$ and $X_\mathrm{H} = 0.1$. \textit{Bottom:} Minimum, maximum and pure-He mass predictions for a fixed luminosity of $\mathrm{log}(L/L_\odot) = 5.144$ and $X_\mathrm{H} = 0.1$.}

\vspace{0.3cm}

\label{tab: compare_works}
\end{table}

\section{Wind models of partially-stripped stars}
\label{sec: powr_wind_pss}

In the previous sections, we saw how structures consisting of a He core and a low-mass H shell resulting from partial envelope stripping can output a total surface luminosity higher compared to their pure-He, fully-stripped counterparts of the same total mass. Similarly, in terms of mass, for a given surface luminosity, these configurations can have a lower mass compared to the pure-He case. In terms of the Eddington parameter $\Gamma_\mathrm{rad}$, which goes as the luminosity-to-mass ratio, they mean the same thing -- an increase in $\Gamma_\mathrm{rad}$. This increase in $\Gamma_\mathrm{rad}$ can have effect on both the structure of such systems as well as impact their wind properties. 

To this end, we run a small sample of hydrodynamically consistent $\texttt{PoWR}^\textsc{hd}$ wind atmosphere models by varying only the luminosity while all other parameters are kept fixed. The fixed quantities are as follows: stellar mass $M_\mathrm{tot} = 10\,M_\odot$, chemical abundance with $X_\mathrm{H} = 0.1$, $Z = 0.008$ with individual mass fractions following solar-scaled ratios from \citet{GS98}, inner boundary temperature $T_\star = 30$ kK defined at Rosseland continuum optical depth of $\tau_\mathrm{Ross,cont}=5$. 

\begin{figure*}[h!]
    \includegraphics[width = \textwidth]{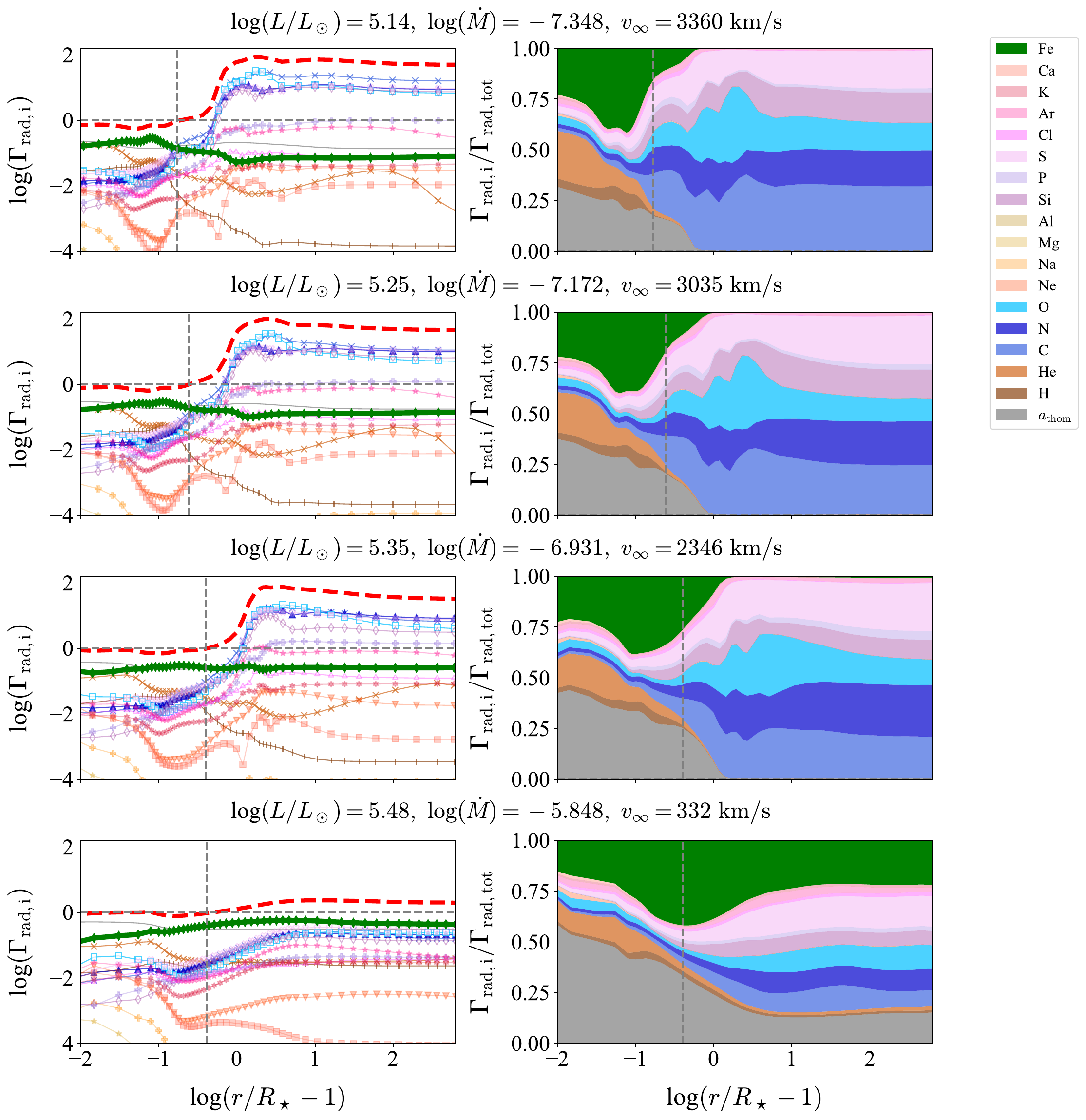}
    \caption{\textit{Left:} Radiative acceleration normalized to gravity as a function of radius for four $\texttt{PoWR}^\textsc{hd}$ models, showing the individual contributions from various elements (coloured) and electron scattering (gray) to the total radiative acceleration (red dashed). \textit{Right:} Radiative acceleration from various elements and electron scattering relative to the total radiative acceleration. The gray dashed vertical line marks the critical point of the hydrodynamic equation. 
    }
    \label{fig: PSS_compare}
\end{figure*}

\begin{figure}
    \includegraphics[width = \columnwidth]{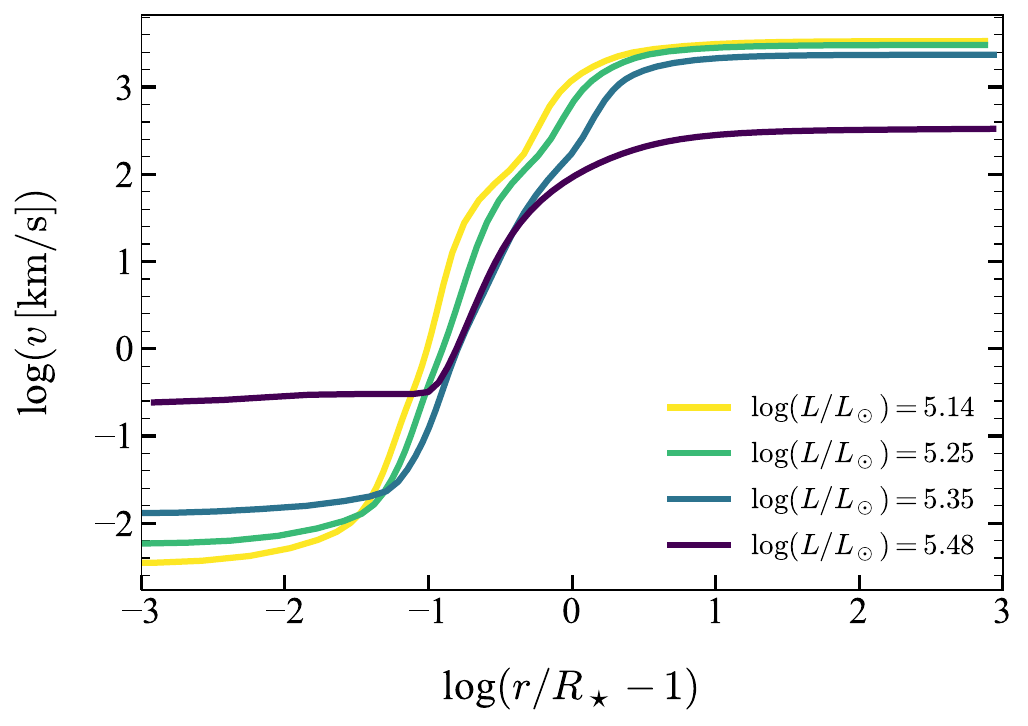}
    \caption{Velocity stratification of the four hydrodynamical wind models from $\mathrm{Fig.}\,\ref{fig: PSS_compare}$.
    }
    \label{fig: PSS_compare_v}
\end{figure}

The inner boundary of our wind simulations lies just below the photosphere, only going down to Rosseland continuum optical depth of 5. However, the impact of the higher  $\Gamma_\mathrm{rad}$ can perhaps already be realised further down in the envelope layers. For example, inflation could occur across opacity zones in near-Eddington situations as seen in 1D structure models or strong convectively-driven turbulence could be triggered as predicted by multi-D radiation hydrodynamic simulations of \citet{Debnath2024}. In this work, we focus on the atmospheric layers where the wind is ultimately launched, while acknowledging that a more sophisticated treatment of turbulence in 1D models, informed by multi-D simulations, is necessary to consistently simulate the deeper layers in our models.

The chosen luminosity values range from log$(L_\star/L_\odot) = 5.14$ to $5.48$, corresponding to the luminosity of a pure-He case to the maximum possible luminosity for a mass of $10\,M_\odot$ and surface $X_\mathrm{H} = 0.1$.

In $\mathrm{Fig.}\,\ref{fig: PSS_compare}$, we present the acceleration stratification of four models with increasing luminosity. On the left column, the total radiative acceleration (shown as a red dashed line) is broken down into its individual contributions from various elements, along with the electron scattering component. On the right column, we show the relative contribution of individual elements and electron scattering to the total radiative acceleration. The grey dashed line indicates the critical point of the hydrodynamic equation, where the flow velocity equals the sonic velocity corrected for turbulent motion and the radiative acceleration balances gravity, that is, the local Eddington parameter crosses unity.

The mass-loss rate increases with luminosity. The higher the luminosity, the higher is the total radiative acceleration in the inner wind which drives a denser wind with higher mass-loss rate and a lower wind terminal velocity \citep[see also wind models in][]{Vink99, Sander2023, Sabhahit_VMS2025}. The changes in the wind  properties are drastic over the range of the tested luminosities. For an increase of 0.34 dex in log($L_\star/L_\odot$), the mass-loss rate increased by 1.5 dex, while the terminal velocity decreased by 1 dex. 

We can further notice drastic changes to the acceleration structure by comparing the individual line contributions from different elements. While iron dominance in driving the wind is evident across all models, iron acceleration increases -- both near the critical point and in the outer super-critical wind regime -- toward the maximum luminosity. At lower luminosities, while iron is still prevalent in the inner wind, other elements such as C, N, O, silicon, sulphur begin to dominate near the critical point. Their dominance continues in the outer wind and completely take over outer wind driving. These trends are consistent with wind models in other parameter regimes \citep[see for example, wind models by][]{Vink2001, Muijres2012, Sander2020b}. 

In $\mathrm{Fig.}\,\ref{fig: PSS_compare_v}$, we show the velocity stratification calculated from our models. The wind velocities too shows a different structure for the model which has the maximum possible luminosity. As mentioned earlier, the terminal velocity of this model is of an order of magnitude lower compared to the pure-He case. 

\begin{figure}
    \includegraphics[width = \columnwidth]{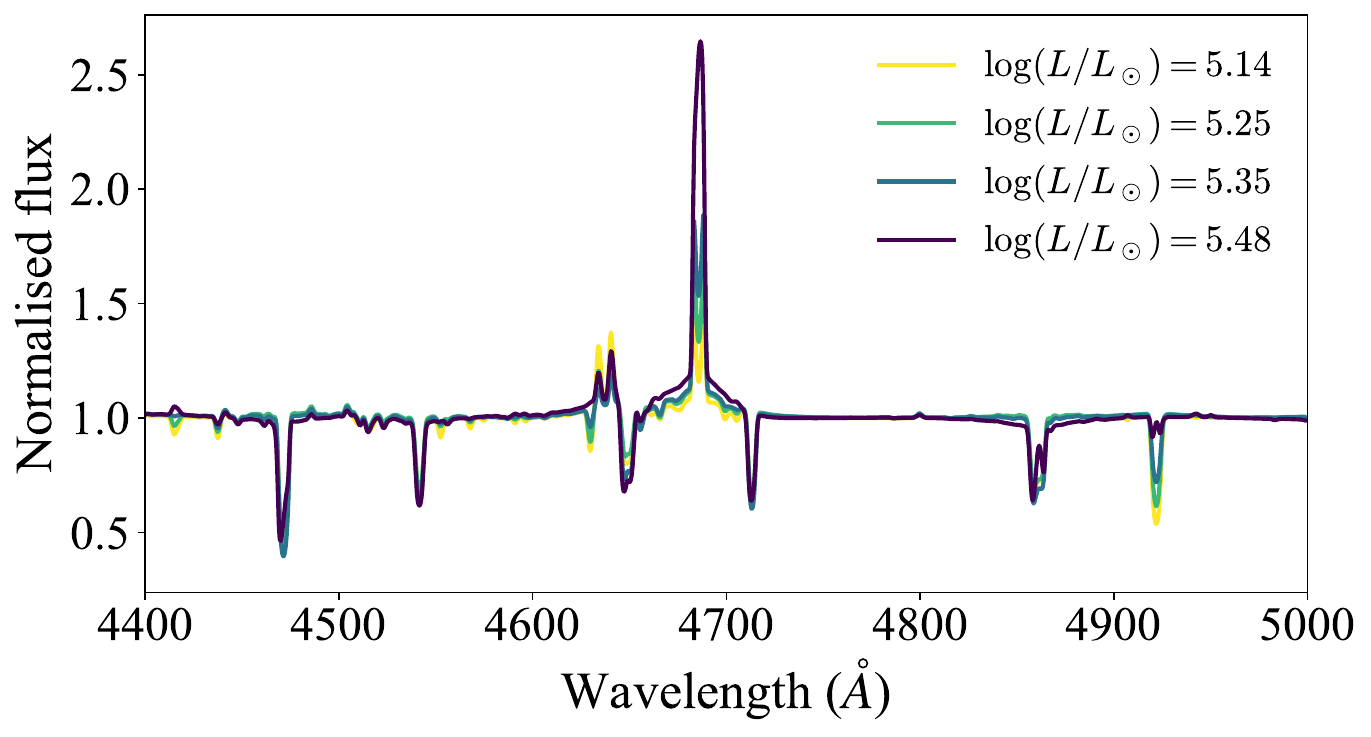}
    \caption{Synthetic normalized spectrum resulting from the four hydrodynamical wind models.}
    \label{fig: PSS_compare_spec}
\end{figure}

\astext{In Fig.\,\ref{fig: PSS_compare_spec}, we show the resulting normalized spectra in the blue region between $4450$ and $4950\,$\AA. For the model with the lowest luminosity, all lines except the \ion{N}{iii} lines defining the ``f'' classification and \ion{He}{ii}\,4686\,\AA\ are in absorption. The \ion{N}{iii} $\lambda$4634-40-42 lines can be in emission even for low wind mass-loss rates due to non-LTE effects, but the very sensitive behaviour of \ion{He}{ii}\,4686\,\AA\ indicates that even for the rather low absolute mass-loss rate of $\log\,(\dot{M}\,[M_\odot/\mathrm{yr}]$) = $-7.35$ there is a notable imprint on the optical spectra. However, only for the higher luminosity \ion{He}{ii}\,4686\,\AA\ appears in pure emission, while for the other three examples it shows a central absorption on top of an emission. This feature is also seen in the so-called Onfp stars \citep{Walborn1973}, but in this classification associated with significant rotation, which is not included in our models. This underlines earlier discussions pointing out that this phenomenon can have more than one origin \citep[e.g.,][]{Vink2009}. Moreover, the apparent double-horn profile could be misinterpreted as a feature of a disk, in particular as partially-stripped stars will in most cases be accompanied by a non-degenerate and often cooler (e.g., Be-type) companion. While such a feature has not been definitively identified in stripped stars (which could be due to observational bias, as we only have a handful of these intermediate stripped objects)}, our results importantly underline that the presence of double-horned \ion{He}{ii} emission in an otherwise cooler-looking spectrum by no means implies the presence of a black hole \citep[cf., e.g.,][]{Casares2014,Janssens2023}.

Finally, we can comment on the future evolution and fate of these systems in light of the drastic increase in the predicted mass-loss rate achievable by partially-stripped stars in their maximum luminosity configuration. 
\gnstext{For a mass-loss rate of log($\dot{M} [M_\odot/\mathrm{yr}]$) $= -5.85$ and an evolutionary timescale of a few $10^5$ years  \citep[see][]{Dutta_Klencki2024}, we estimate a total mass loss of $0.1-0.5\,M_\odot$. While this might seem negligible, some stripped-stars could have envelope masses of the same order, meaning that mass loss could completely remove their envelope, forming cWRs or low-mass stripped He stars. However, this is an optimistic estimate of the total mass lost. First, the mass of $10\,M_\odot$ is on the higher end for typical stripped stars. Additionally, not all observed stripped stars will have the specific maximum luminosity configuration, meaning they will typically have lower mass-loss rates and higher terminal velocities. Indeed, the stars analysed by \citet{Ramachandran2024} show lower mass-loss rates and higher velocities of the order 1000 km/s. However, the model stellar parameters chosen here differ from the empirically determined stellar parameters in \citet{Ramachandran2024}, and a strict comparison of the wind properties requires spectral fits using hydrodynamic models with similar stellar parameters.} 

The higher possible mass-loss rates, however, is relevant given that the generally low ejecta masses from SNe Ibc studies \citep{Shigeyama1990, Drout2011, Lyman2016} might suggest binary stripping. On the other hand, if these partially-stripped stars were progenitors of SNe Ibc, they should be bright enough to be detectable in the optical, but so far, they have not been observed. This could further hint in the direction of cWRs as SNe Ibc progenitors \citep{Yoon2012_ibc, Groh2013}, given that full stripping might occur due to the higher possible mass-loss rate during the partially-stripped star phase.

While we presented a small sample of hydrodynamically consistent wind models, using stellar parameters informed by partially-stripped structures, a more detailed and comprehensive study of the wind properties across the full relevant parameter space is needed in the future, as mass loss during the stripped phase is crucial in the context of whether these objects ultimately explode as Type IIb or Ibc stripped-envelope SNe \citep{Vink2017, Gilkis2019}.

\sisetup{group-separator = {}}
\renewcommand{\arraystretch}{1.3}
\begin{table*}[h!]
\centering
\footnotesize
\caption{Empirically determined luminosity, surface H mass fraction, and spectroscopic mass of four partially-stripped systems in the MCs. }
\begin{tabular}{|p{2.5cm}|p{2.2cm}|p{2.cm}|p{2.cm}|p{2.cm}|}
\hline
 & SMCSGS-FS 69 & 2dFS 163 & 2dFS 2553 & \text{Sk~$-71^\circ~35$}  \\ 
\hline
log$(L/L_\odot)$   & $4.7^{+0.1}_{-0.1}$ & $4.75^{+0.1}_{-0.1}$ & $4.91^{+0.1}_{-0.1}$ & $4.93^{+0.1}_{-0.1}$ \\
$X_\mathrm{H}$ [mass fr.] & $0.59^{+0.1}_{-0.1}$ & $0.33^{+0.1}_{-0.05}$ & $0.6^{+0.1}_{-0.05}$ & $0.7^{+0.04}_{-0.1}$ \\
$M_\mathrm{spec}$ [$M_\odot$] & $2.8^{+1.5}_{-0.8}$ & $3.96^{+1.8}_{-1.5}$ & $6.86^{+3.3}_{-2.8}$ & $7.8^{+3.6}_{-3}$ \\
\hline
$M_\mathrm{max}$ [$M_\odot$] & 18.08 & 13.063  & 22.177 & 25.672 \\
$M_\mathrm{He}$ [$M_\odot$] & 6.217 & 6.538 & 7.721 & 7.887 \\
$M_\mathrm{min}$ [$M_\odot$] & 4.132 & 3.871 & 5.465 & 5.769 \\
\hline
\end{tabular}
\tablefoot{The empirical stellar parameters are taken from \citet{Ramachandran2023} and \citet{Ramachandran2024}. We compare the spectroscopic masses with the maximum, pure-He, and minimum masses predicted by the MLRs presented in this work.}
\vspace{0.3cm}

\label{tab: stripped_stars}
\end{table*}

\section{Implications for observed partially-stripped stars}
\label{sec: discussion}

\gnstext{We now discuss the implications of our MLRs on the evolutionary status of observed partially-stripped stars. \citet{Ramachandran2023, Ramachandran2024} recently analysed the spectroscopy of four partially-stripped stars in the MCs, three from the X-Shooting ULLYSES (XShootU) sample and one from the Fiber Large Array Multi-Element Spectrograph
(FLAMES) on the Very Large Telescope (VLT). 
The empirically estimated luminosity, surface $X_\mathrm{H}$, and spectroscopic mass of the four objects are shown in $\mathrm{Table}\,\ref{tab: stripped_stars}$. Using this luminosity and surface $X_\mathrm{H}$, we list the three different mass estimates according to our MLRs. While the maximum mass chemical homogeneous evolution scenario has little relevance for the partially-stripped objects discussed here, we still provide the mass estimate for comparison purposes.} 

In the following paragraphs, we show that our minimum and maximum MLRs satisfactorily bracket the observed masses and luminosities of partially stripped systems, with certain systems offering predictive power regarding their internal structure. Our structure models, however, predict surface temperatures lower than empirical estimates, which, as discussed at the end of $\mathrm{Sec.}\,\ref{sec: L_vs_slope}$, could be an artefact of uncertainties in how inflation is treated in these near-Eddington-limit models. We therefore place less diagnostic weight on temperature than compared to luminosity and mass. 

\gnstext{For the object SMCSGS-FS 69, we note that the spectroscopic mass is close to the minimum mass (within error bars) and well below the pure-He mass and the chemical homogeneous mass. \citet{Ramachandran2023} already noted that the case A mass transfer scenario fails to reproduce the spectroscopic mass and $\mathrm{log}\,g$ of this object, and in the mass-luminosity space it sits comfortably above the pure-He MLR from \citet{Graf2011}. Assuming that the minimum mass is close to and within the error bar of the spectroscopic mass, we estimate an internal slope value of $s\approx 2.56$ using $\mathrm{Eq.}\,\ref{eq: L_giv_M_X_slope}$. The low value of $s$ suggests that the prior evolution of this star did not pass through an advanced core-He burning phase, for example as an RSG. Instead, the shallow slope favours partial stripping via case-B binary channel} either very early-on during core-He burning or during the HG expansion phase. If SMCSGS-FS 69 had undergone wind mass-loss stripping as an advanced core-He burning object, the resulting H slope would be significantly higher than the value of 2.56 that gives the current mass. 

It is important to note here that while a shallow H slope is unlikely to result from the single-star wind-stripping channel and strongly hints at case-B binary mass transfer, a steeper slope can result from both single-star wind-stripping and binary channels. In the binary channel, a high $\alpha_\mathrm{sc}$ value could potentially steepen the slope already during the HG crossing phase, before the mass transfer occurs (in our evolution testing, most of the $\alpha_\mathrm{sc} = 100$ models do this). Therefore, for steeper slopes, given the uncertainties in mass and internal mixing processes in massive stars, it may not be possible to unambiguously determine the evolutionary pathway leading to the stripped star phase.

\gnstext{A similar analysis can be performed for 2dFS 163, which has a spectroscopic mass close to its minimum mass. Within the framework of the structure grid we have built, the estimated slope is $s\approx 2$, which is similar to the first object. This hints at a prior evolution involving binary stripping during the early core-He burning or HG expansion phase. However, we would like to note that 2dFS 163 has a relatively low $X_\mathrm{H} \sim 0.33$ compared to the other cool partially-stripped objects discussed here.  \citet{Ramachandran2024} hinted towards this object being a post-core He burning giant undergoing (re-)expansion upon core-He exhaustion \citep{Laplace2020}. We note that our grid is built to resemble a typical partially-stripped, post-mass transfer configuration with a He core and an H shell, and does not take into account the scenario of post-He burning objects, which features both a He shell and an H shell \citep{Laplace2020}. Such a configuration would technically correspond to an even lower mass (compared to the minimum we predict here with the typical partially-stripped structure) for a given luminosity and surface $X_\mathrm{H}$. 
While also incorporating post-He burning structures might require future efforts, we note that from an evolutionary timescale perspective these post-He burning objects evolve on thermal timescales. In contrast, stars can spend up to $10\%$ of the core-He burning lifetime in a typical partially-stripped structure \citep{Dutta_Klencki2024}. Therefore, most observed partially-stripped stars are likely still early core-He-burning objects.}

The third object, 2dFS 2553, has a spectroscopic mass approximately halfway between the minimum mass and the pure-He mass. \citet{Ramachandran2024} also provide orbital masses from orbital analysis, placing the $M_\mathrm{orb} = 7.5^{+5.8}_{-2.3}\,M\odot$ very close to its pure-He mass. The final object, \text{Sk~$-71^\circ~35$}, also has a spectroscopic mass near the pure-He mass. Despite their masses being comparable to their pure-He masses, their surface $X_\mathrm{H}$ values remain high ($\sim 0.6-0.7$). As seen from $\mathrm{Fig.}\,\ref{fig: logL_vs_slope}$, such a scenario is possible, where the surface $X_\mathrm{H}$ value remains high while the mass is comparable to the pure-He mass. While we cannot use our minimum and maximum MLRs to predict masses in such scenarios, we can interpolate between our structure grid to predict the slope values that give pure-He masses for the given surface $X_\mathrm{H}$. 

Given the non-monotonic behaviour of luminosity with slope, two different slope values -- a low and a high value -- can give the pure-He mass. For example, for 2dFS 2553, we find $s\approx 1.25$ and $s\gtrsim 6$ slope values. A note of caution is needed here, as the spectroscopic mass has large uncertainties spanning the range from the minimum to the pure-He mass, meaning the entire range of slope values $s \gtrsim 1.25$ is possible. We therefore can not constrain the prior evolution as there are multiple evolution channels which could result in these values.
While the low $s$ values can only be achieved through binary stripping, the high values can result from both binary stripping -- for example, from nuclear timescale mass transfer \citep[see e.g.,][]{Klencki2022} -- as well as from wind mass-loss stripping during an advanced core-He burning phase.

\section{Overview and conclusions}
\label{sec: conclusions}

In this work we present a large grid of structure models aimed at revisiting the mass-luminosity relations for massive stars with the inclusion of partially-stripped structures. A typical partially-stripped structure consists of a He-rich core with a low-mass H shell on top, having lost the rest of their envelope, either through  single-star wind stripping or binary mass transfer. This leaves behind a low-mass, H-depleted envelope burning H in a shell around a He-burning core.

We first detail our method to construct a grid of structure models with four different variable parameters: the total mass $M_\mathrm{tot}$, the total metal mass fraction $Z$ taking values, surface $X_\mathrm{H}$, and the slope $s$ of the H profile which depletes from the surface value to 0 in the He core. The slope $s = dX/dQ$ is defined similar to \citet{Abel2018, Abel2019}, which uses a normalised mass-coordinate $Q$ going from 0 at the centre to 1 at the location where the extrapolated H mass fraction value equals its initial value. For $0 < s < \infty$, the grid covers the following range: $1\leq M_\mathrm{tot}/M_\odot \leq 18$ and $0.01\leq X_\mathrm{H} \leq 0.7$. Additional structure models are run corresponding to the fully chemically homogeneous case where the H in the entire star is the same as the surface ($s=0$), and pure-He models ($s=\infty$). For these $s = 0$ and $s = \infty$ models, the grid extends to $1\leq M_\mathrm{tot}/M_\odot \leq 40$ with the same range of $X_\mathrm{H}$. All models are calculated for two metallicities, $Z = 0.008$ and $Z = 0.004$.

The main reason for revisiting the MLRs is that partially stripped structures, consisting of both a He-burning core and an H-burning shell, can violate the simple $L \sim \mu^4 M^3$ scaling derived from homology relations.
These systems can be over-luminous compared to their pure-He counterparts of the same mass, sometimes by more than a factor of two. This is due to the H shell dominating the total luminosity over the He-core luminosity. 

We show that for a given mass, the luminosity is a non-monotonic function of the slope as $s$ increases from 0 to $\infty$. For all values of $s$, the $s=0$ chemically homogeneous case has the least luminosity. The luminosity then rapidly increases, reaching a maximum at some slope $0 < s < \infty$. Beyond this slope, the luminosity decreases and asymptotically approaches the $s=\infty$ pure-He case. \gnstext{This non-trivial behaviour of luminosity is due to the competing effects of the H-shell luminosity disproportionately dominating the total luminosity budget and the effects of mean molecular weight and} opacity in the H envelope.

We then provide convenient MLR fits to predict the minimum ($s = 0$), the maximum $( 0 < s < \infty)$, and the pure-He luminosity ($s = \infty$) for a given total mass and surface $X_\mathrm{H}$. The relations can be numerically inverted to get the minimum, the maximum, and the pure-He masses for a given surface luminosity and surface $X_\mathrm{H}$. 

The higher luminosity achievable from such partially-stripped structures for a given mass allows them to easily approach their local Eddington limit. We demonstrated how the increased Eddington parameters in their atmospheres impact wind properties, leading to drastic changes in acceleration and velocity structure, as well as an increase in mass-loss rates.

\gnstext{We also briefly discuss the implications of our MLRs for the observed partially-stripped stars from \citet{Ramachandran2023, Ramachandran2024}. For SMCSGS-FS 69 and 2dFS 163, the spectroscopic mass is very close to the minimum mass predicted by our MLRs for its given luminosity and surface $X_\mathrm{H}$. Based on our framework, we predict an internal structure with an H-profile slope close to 2. The shallow slope favours binary stripping, occurring either during HG expansion or early core-He burning,} rather than wind-driven mass-loss stripping from a more advanced core-He burning stage. However, 2dFS 163 has a low(er) surface $X_\mathrm{H}$, and could be a post-He burning (re-)expanding giant. Future efforts are needed to incorporate such post-He burning giant structures.  For 2dFS 2553 and \text{Sk~$-71^\circ~35$}, the situation is less clear due to larger error bars, and we provide a range of possible slope values that their structure could potentially have. Furthermore, while our minimum and maximum MLRs bracket the spectroscopic masses of observed partially-stripped systems, we would like to note that our models predict lower surface temperatures than empirical estimates. This discrepancy may arise from how inflation is treated in the envelopes of such systems, and the associated uncertainties in the predicted radius and temperature. Consequently, given these temperature uncertainties, our comparison with observed partially stripped systems relies primarily on mass and luminosity.

Given the low-mass envelopes of partially-stripped stars, the drastic change in mass loss at higher Eddington parameters achievable in such systems, and their significance for future evolution and supernova type they can explode as, a detailed investigation of partially-stripped star winds across the relevant parameter range (including the mass-loss rate (MLR) results presented here) is needed in the future.

\begin{acknowledgements}
 We thank the anonymous referee for constructive comments that helped improve the paper. GNS ad JSV are supported by STFC funding under grant number ST/Y001338/1. AACS and VR are supported by the German
    \emph{Deut\-sche For\-schungs\-ge\-mein\-schaft, DFG\/} in the form of an Emmy Noether Research Group -- Project-ID 445674056 (SA4064/1-1, PI Sander). AACS further acknowledges financial support by the Federal Ministry for Economic Affairs and Climate Action (BMWK) via the German Aerospace Center (Deutsches Zentrum f\"ur Luft- und Raumfahrt, DLR) grant 50 OR 2503 (PI: Sander). This project was co-funded by the European Union (Project 101183150 - OCEANS).
\end{acknowledgements}

\bibliographystyle{aa}
\bibliography{References.bib}

\appendix

\section{Testing CNO abundances at ZAMS against CNO cycle equilibrium values}
\label{appendix: CNO_testing}

Here we test the effects of our choice of CNO abundances on the luminosity profile of our models and ultimately the minimum and maximum luminosities predicted by our structure models. The models presented in the work so far fixed the CNO abundance to the values at ZAMS. This is an assumption we make when building our structure model, but here we test this assumption against structure models built with more realistic CNO abundances. This is especially relevant for the individual CNO abundances, as C and O depletes in  N due to the CNO cycle which is the dominant H burning process that occurs in the cores of massive stars.

\begin{figure}[h]
    \includegraphics[width = \columnwidth]{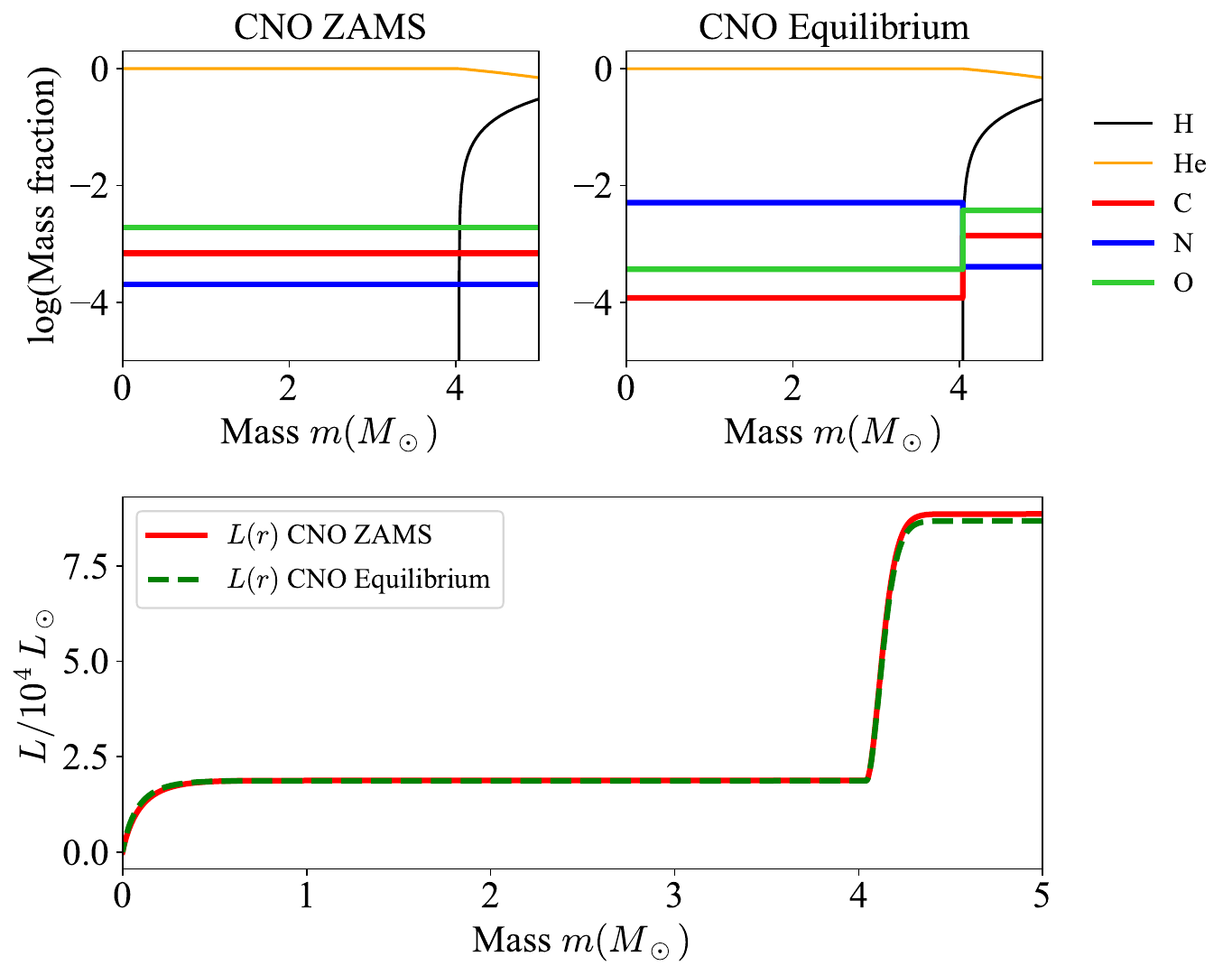}
    \caption{\textit{Top:} Internal chemical profiles showcasing two different CNO abundances: ZAMS values and equilibrium values reflecting higher N in expense of C and O in the He-rich core. \textit{Bottom:} Luminosity profiles of the two models shown in the top sub-plots. }
    \label{fig: CNO_compare}
\end{figure}

In $\mathrm{Fig.}\,\ref{fig: CNO_compare}$, we show the internal chemical profiles showcasing two different CNO abundances. The left sub-plot shows constant CNO abundances fixed to their ZAMS values throughout the model (this is the exact model as in $\mathrm{Fig.}\,\ref{fig: chem_profile}$). The right sub-plot shows CNO equilibrium values that reflects more realistic CNO equilibrium abundance values in the He-rich core, i.e., N-enhanced and CO-poor material. For the actual CNO values in the right sub-plot, we take the average of the CNO abundances obtained from our evolution models in $\mathrm{Sect.}\ref{sec: slope_values_evo}$ at the end of their MS. The four main parameters $M_\mathrm{tot}$, total $Z$, $X_\mathrm{H}$ and $s$ are fixed to the same value as in $\mathrm{Fig.}\,\ref{fig: chem_profile}$. The only change between the models is the individual CNO abundances in the core. The bottom sub-plot in $\mathrm{Fig.}\,\ref{fig: CNO_compare}$ shows the local luminosity stratification of the two models. The difference between the surface luminosity of these two models is roughly 0.01 dex. 

\begin{figure}[h]
    \includegraphics[width = \columnwidth]{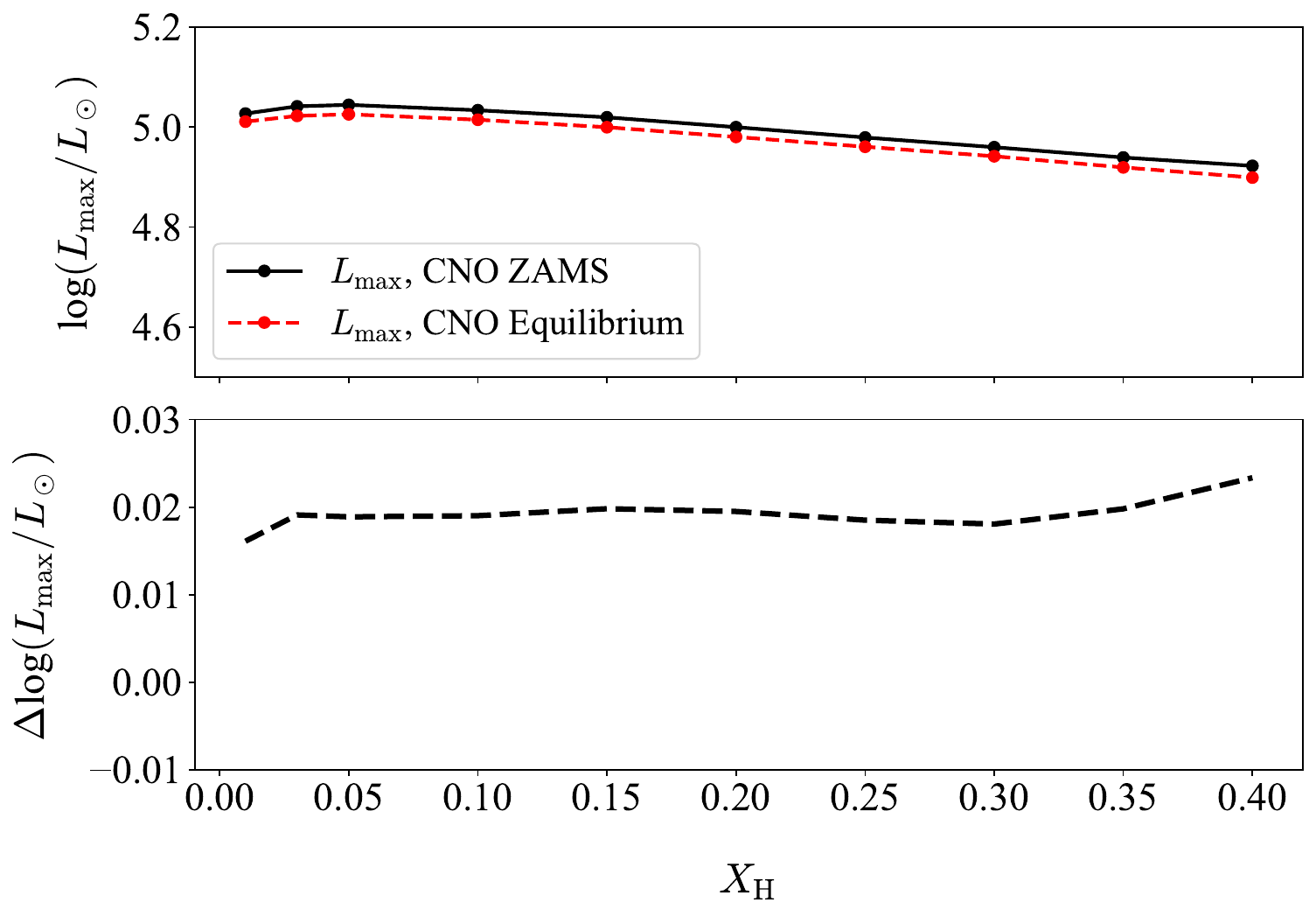}
    \caption{\textit{Top:} Maximum luminosity predicted for $M_\mathrm{tot} = 5\,M_\odot$ using two different CNO abundances: ZAMS and equilibrium values. \textit{Bottom:} The difference in dex between the above predicted maximum luminosities. }
    \label{fig: compare_L_diff}
\end{figure}

To perform a systematic study, we run an additional smaller grid of structure models using the above CNO equilibrium chemical profile. The total mass of this smaller grid is fixed to $M_\mathrm{tot} = 5\,M_\odot$, while the $X_\mathrm{H}$ and $s$ are varied, and can be directly compared to the $M_\mathrm{tot} = 5\,M_\odot$ models shown in $\mathrm{Fig.}\,\ref{fig: logL_vs_slope}$ which CNO abundances of ZAMS. In $\mathrm{Fig.}\,\ref{fig: compare_L_diff}$, we show the maximum luminosity as a function of $X_\mathrm{H}$ using the two different CNO profiles. In the bottom sub-plot, we also show the absolute difference between the luminosities which is roughly 0.02 dex.

\section{Variation of luminosity with slope for different masses}
\label{appendix: diff_mass}
\begin{figure*}
    \includegraphics[width = \textwidth]{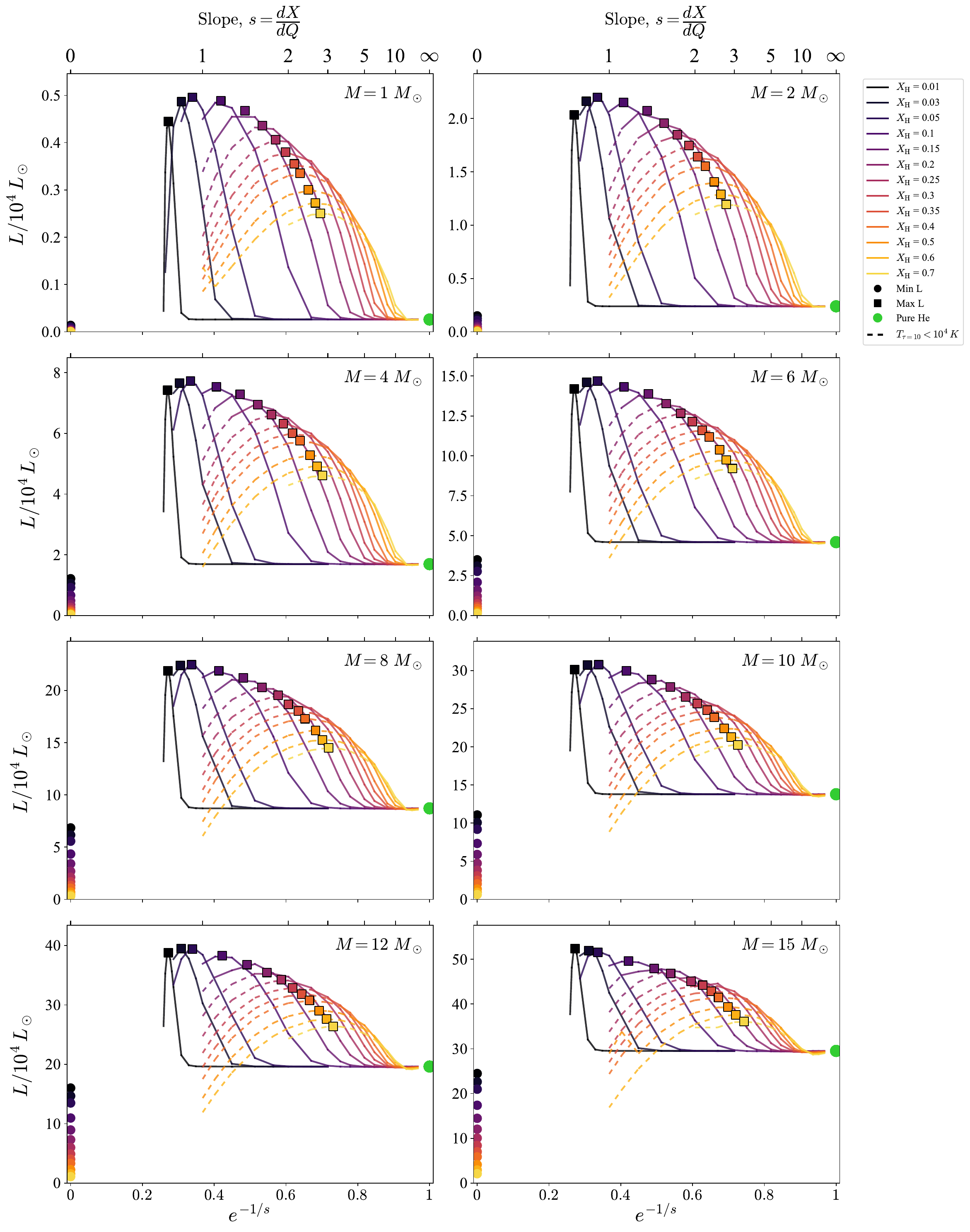}
    \caption{Same as $\mathrm{Fig.}\,\ref{fig: logL_vs_slope}$, but for different $M_\mathrm{tot}$ values selected from our grid. The coloured dots and squares correspond to minimum and maximum luminosities, and the green dot corresponds to the pure-He case retaining their meaning from $\mathrm{Fig.}\,\ref{fig: logL_vs_slope}$.}
    \label{fig: logL_vs_slope_diff_mass}
\end{figure*}

In $\mathrm{Fig.}\,\ref{fig: logL_vs_slope_diff_mass}$, we show the variation of luminosity with slope $s$ for eight select model sequences with different total masses.

\section{Minimum and maximum luminosities at $Z = 0.004$}
\label{appendix: SMC_Z}
\begin{figure*}[h]
    \includegraphics[width = \textwidth]{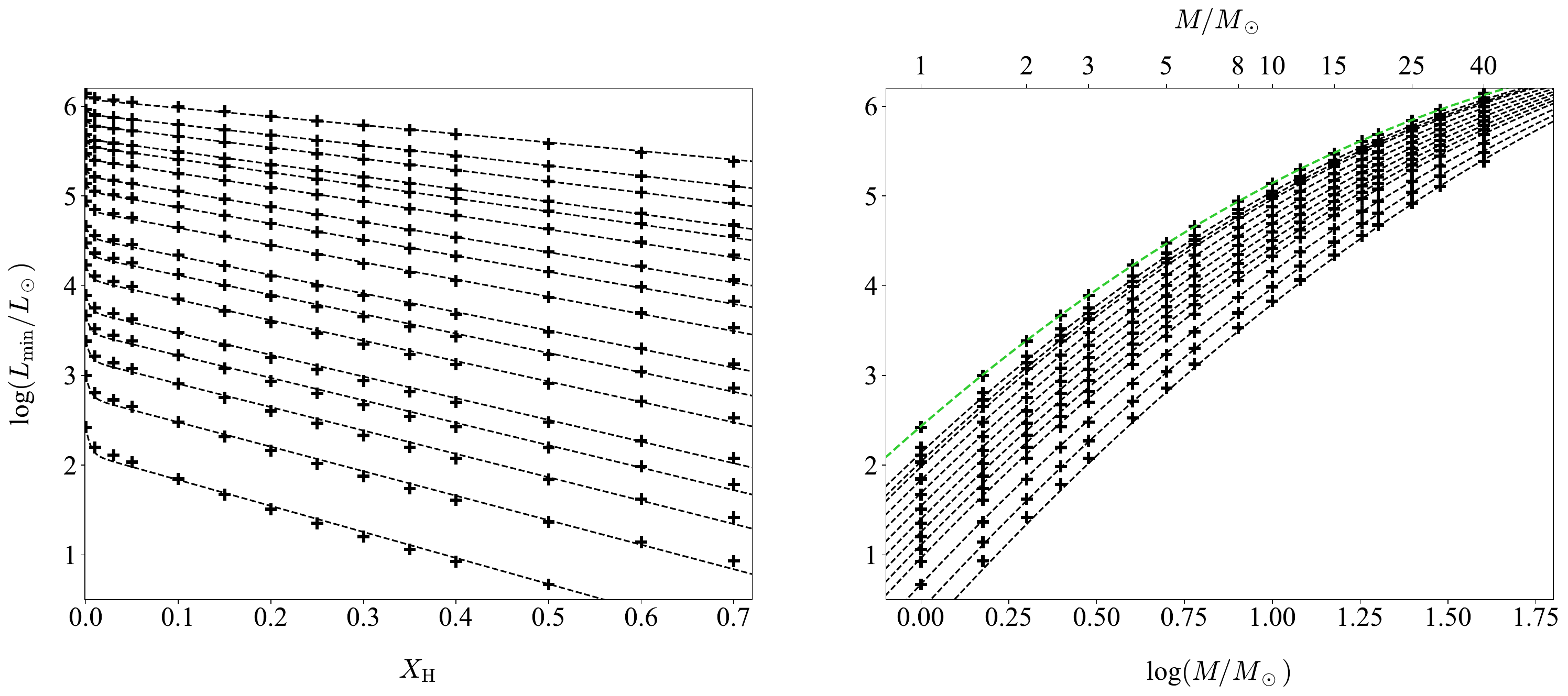}
    \caption{Same as $\mathrm{Fig.}\,\ref{fig: Min_L_given_M_X}$, but for $Z = 0.004$.
}
    \label{fig: Min_L_given_M_X_SMC}
\end{figure*}

\begin{figure*}[h]
    \includegraphics[width = \textwidth]{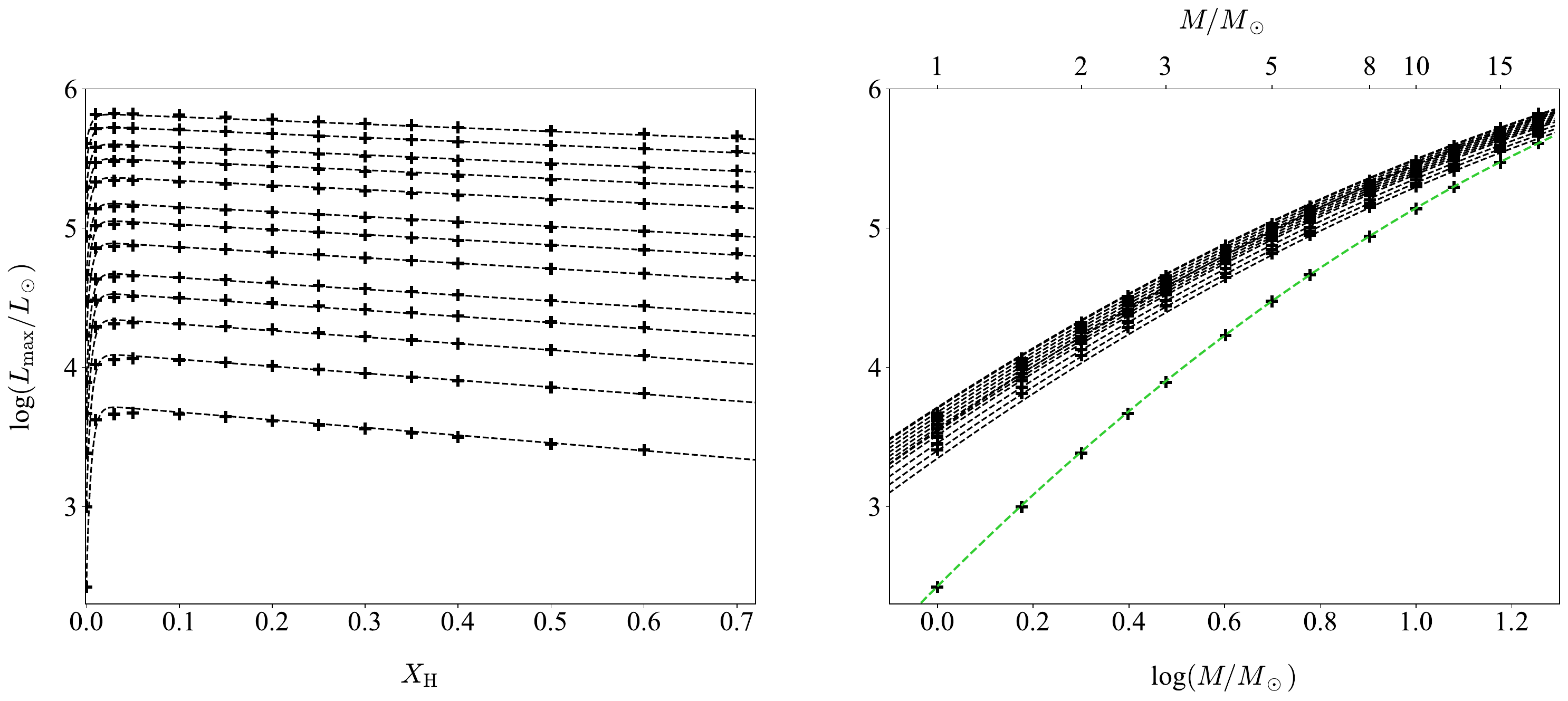}
    \caption{Same as $\mathrm{Fig.}\,\ref{fig: Max_L_given_M_X}$, but for $Z = 0.004$.
    }
    \label{fig: Max_L_given_M_X_SMC}
\end{figure*}

In $\mathrm{Figs.}\,\ref{fig: Min_L_given_M_X_SMC}$ and $\ref{fig: Max_L_given_M_X_SMC}$, we show the minimum and maximum luminosities for given total mass and surface $X_\mathrm{H}$ at total metal mass fraction of $Z = 0.004$. The black dashed lines are predictions from our best-fit MLRs from $\mathrm{Eq.}\,\ref{eq: L_giv_M_X}$, with the coefficients $F_1$ to $F_{10}$ provided in $\mathrm{Table}\,\ref{tab:parameters}$.

\sisetup{group-separator = {}}
\renewcommand{\arraystretch}{1.4}

\begin{table}
\centering
\caption{Values of coefficients $F_1$ to $F_{10}$ to predict $L_\mathrm{min}$ and $L_\mathrm{max}$ for $Z = 0.008$ and $0.004$.}
\begin{tabular}{|c| c |c | c| c|}
\hline
 & \multicolumn{2}{c|}{$Z = 0.008$} & \multicolumn{2}{c|}{$Z = 0.004$} \\ \hline
 & {$L_\mathrm{min}$} & {$L_\mathrm{max}$} & {$L_\mathrm{min}$} & {$L_\mathrm{max}$} \\ \hline
$F_1$ & 2.053491 & 3.751088 & 2.125432 & 3.733297 \\
$F_2$ & 3.790927 & 2.209607 & 3.689468 & 2.198926 \\
$F_3$ & $-0.802070$ & $-0.453056$ & $-0.763519$ & $-0.424813$ \\
$F_4$ & $-2.976704$ & $-0.520778$ & $-2.900812$ & $-0.552451$ \\
$F_5$ & 0.965973 & 0.245808 & 0.934060 & 0.309716 \\
$F_6$ & 0.185089 & $-0.016714$ & 0.173159 & $-0.060483$ \\
$F_7$ & 0.369268 & $-1.329120$ & 0.308744 & $-1.305613$ \\
$F_8$ & $-0.374144$ & 1.228870 & $-0.307890$ & 1.228668 \\
$F_9$ & 0.105449 & $-0.262928$ & 0.090761 & $-0.286156$ \\
$F_{10}$ & 0.005 & 0.005 & 0.005 & 0.005 \\
\hline
\end{tabular}%

\vspace{0.3cm}

\label{tab:parameters}
\end{table}

\section{Fit relations to predict slope which gives $L_\mathrm{max}$}
\label{appendix: slope_fits}

\sisetup{group-separator = {}}
\renewcommand{\arraystretch}{1.4}

\begin{table}
\centering
\caption{Values of coefficients $F_1$ to $F_9$ to predict the value of slope $s$ which maximises the luminosity for $Z = 0.008$ and $Z = 0.004$.}
\begin{tabular}{|c| c |c |}
\hline
 & {$Z=0.008$} & {$Z=0.004$} \\ \hline
$F_1$ & 0.698967 & 0.709244 \\
$F_2$ & $-0.025170$ & 0.007519 \\
$F_3$ & 0.003576 & $-0.020923$ \\
$F_4$ & 5.017684 & 4.636537 \\
$F_5$ & $-1.125765$ & $-1.631714$ \\
$F_6$ & 1.362459 & 1.744423 \\
$F_7$ & $-2.995227$ & $-2.711640$ \\
$F_8$ & 1.177010 & 2.088682 \\
$F_9$ & $-0.692827$ & $-1.369483$ \\
\hline
\end{tabular}%
\vspace{0.3cm}

\label{tab: slope_coeffs}
\end{table}

\begin{figure*}[h]
    \includegraphics[width = \textwidth]{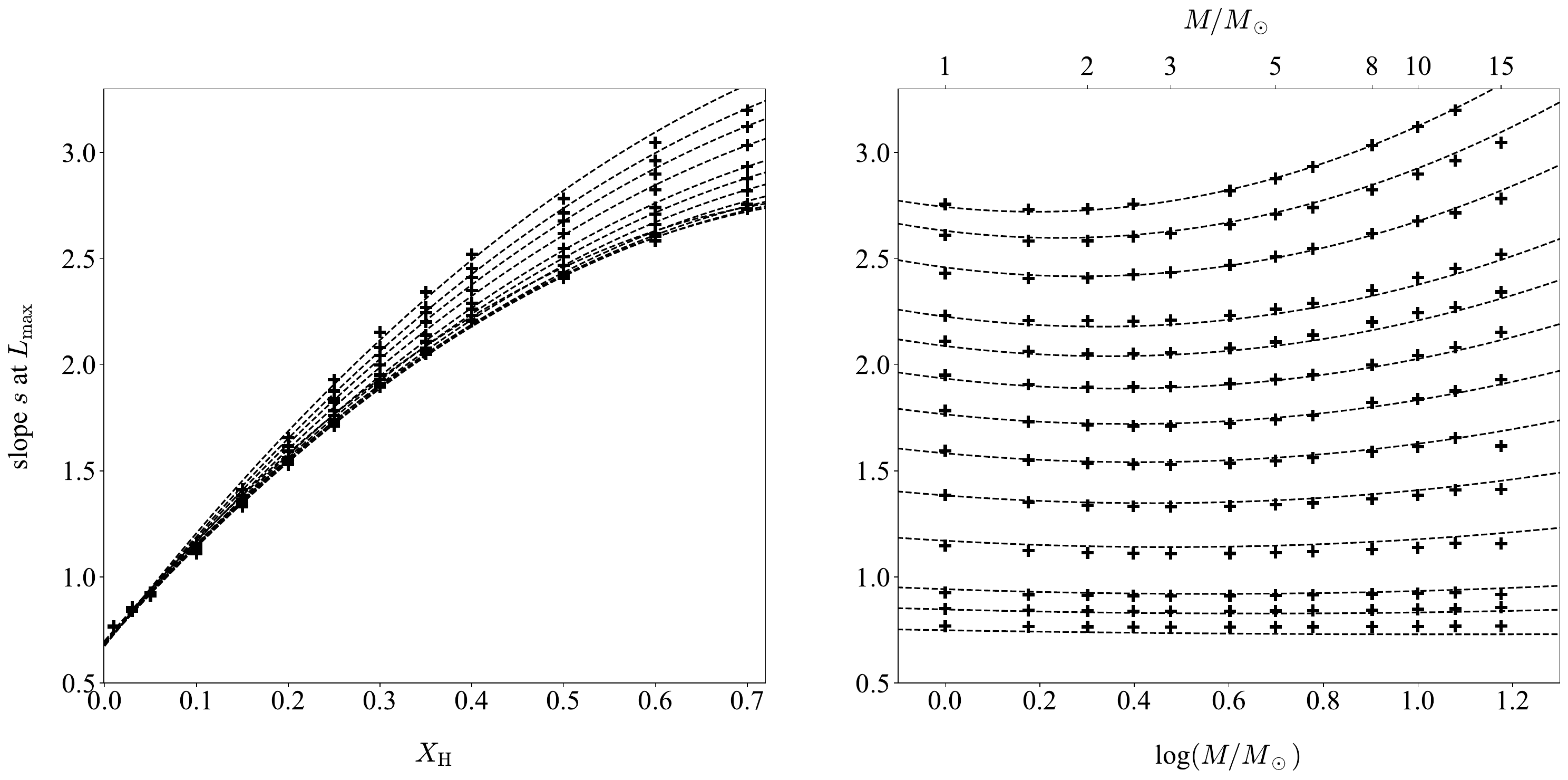}
    \caption{Slope $s$ which maximises the luminosity for given total mass and surface $X_\mathrm{H}$ for $Z = 0.008$. \textit{Left:} $s$ as a function of surface $X_\mathrm{H}$ for total masses ranging from $1\,M_\odot$ to $18\,M_\odot$.   \textit{Right:} $s$ as a function of total mass for $X_\mathrm{H}$ values ranging from 0.01 to 0.7. The plus signs indicate estimates from our structure models, while the dashed black lines represent our best-fit formula  from Eq. \ref{eq: L_giv_M_X_slope}}

    \label{fig: Min_L_given_M_X_slope}
\end{figure*}

\begin{figure*}[h]
    \includegraphics[width = \textwidth]{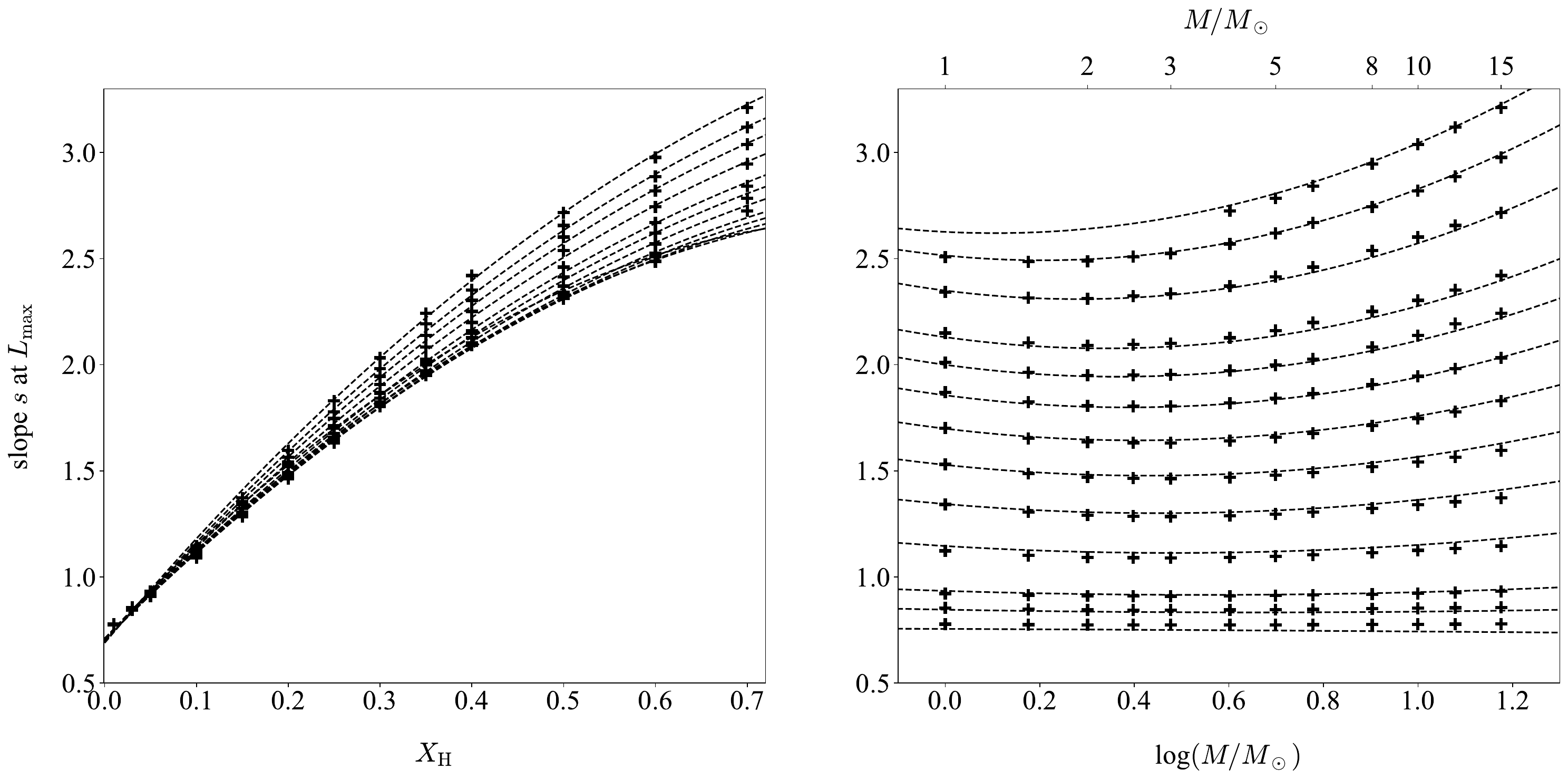}
    \caption{Same as $\mathrm{Fig.}\,\ref{fig: Min_L_given_M_X_slope}$, but for $Z = 0.004$. 
    }
    \label{fig: Max_L_given_M_X_SMC_slope}
\end{figure*}

\noindent
Here we provide best-fit relations to predict the value of slope $s$ which maximises the luminosity for a given total mass and surface $X_\mathrm{H}$. As for the functional form, we  use quadratic terms in both $M_\mathrm{tot}$, and in $X_\mathrm{H}$. The relations are applicable in the ranges $1 \leq M_\mathrm{tot}/M_\odot \leq 18$ and $0.01 \leq X_\mathrm{H} \leq 0.7$. The relations are as follows:
\vspace{0.5cm}
\begin{equation}
\begin{split}
s & = \,\,  F_1 + F_2 \log\left(M/M_\odot\right) + F_3 \log\left(M/M_\odot\right)^2 \\[8pt] 
& + \Bigg[F_4 + F_5 \log\left(M/M_\odot\right) + F_6 \log\left(M/M_\odot\right)^2 \Bigg] \;X_\mathrm{H} \\[8pt] 
& + \Bigg[F_7 + F_8 \log\left(M/M_\odot\right) + F_9 \log\left(M/M_\odot\right)^2 \Bigg]\; X_\mathrm{H}^2
\end{split}
\label{eq: L_giv_M_X_slope}
\end{equation}
where two separate set of coefficients $F_1$ to $F_9$ are provided in $\mathrm{Table}\,\ref{tab: slope_coeffs}$ corresponding to $Z = 0.008$ and $Z = 0.004$. The slope value $s$ which maximises the luminosity (or minimises the mass) estimated by our structure models and our best-fit predictions are shown in $\mathrm{Figs.}\,\ref{fig: Min_L_given_M_X_slope}$ and $\ref{fig: Max_L_given_M_X_SMC_slope}$.

\section{Python script for the mass-luminosity relations}
\label{appendix: python_script}

A python script and an online calculator to calculate minimum, maximum and pure-He mass-luminosity relations are available at \href{https://mdot-com.github.io/scripts/}{https://mdot-com.github.io/scripts/}.

Here we provide a brief description of how the Python script works. There are two main calculator modes in the Python script. First is the luminosity calculator, which calculates the minimum, maximum, and pure-He luminosities for a given input mass, surface $X_\mathrm{H}$, and surface metal mass fraction $Z$. The second is the mass calculator, which calculates the minimum, maximum, and pure-He masses for a given input luminosity, surface $X_\mathrm{H}$, and $Z$. The script also outputs the value of the H-profile slope $s$ that gives the minimum, maximum, and pure-He masses or luminosities.

Several warning messages are included to inform the user about the range of our synthetic model grid and any errors regarding the input values. For example, if the output minimum mass (corresponding to a partially stripped configuration with $0 < s < \infty$) is beyond the tested range in this work of $1 < M_\mathrm{tot}/M_\odot < 18$, then a warning message is printed. Or if the user inputs values of $X$ and $Z$ such that $X + Z > 1$, then an error is printed.

For $Z$ values other than the tested values of $Z = 0.008$ and $0.004$, interpolation or extrapolation is performed when calculating the luminosities or masses using these two values. A warning is also printed indicating whether the values are interpolated or extrapolated.

\end{document}